# Inhomogeneous superconductivity in $Lu_xZr_{1-x}B_{12}$ dodecaborides with dynamic charge stripes

A. Azarevich[1], A. Bogach[1], V. Glushkov[1], S. Demishev[1], A. Khoroshilov[1], K. Krasikov[1], V. Voronov[1], N. Shitsevalova[2], V. Filipov[2], S. Gabáni[3], K. Flachbart[3], A. Kuznetsov[4], S. Gavrilkin[5], K. Mitsen[5], S. J. Blundell[6], N. Sluchanko[1]

[1]*Prokhorov General Physics Institute of the Russian Academy of Sciences, Vavilov str. 38, Moscow 119991, Russia*

[2]*Frantsevich Institute for Problems of Materials Science, National Academy of Sciences of Ukraine, 03680 Kyiv, Ukraine*

[3]*Institute of Experimental Physics SAS, Watsonova 47, 04001 Košice, Slovakia*

[4]*National Research Nuclear University MEPhI, 31, Kashirskoe Shosse, 115409 Moscow, Russia*

[5]*Lebedev Physical Institute of RAS, 53 Leninskiy Avenue, 119991 Moscow, Russia*

[6]*Department of Physics, University of Oxford, Clarendon Laboratory, Parks Road, Oxford OX1 3PU, United Kingdom*

**Abstract.**

We have studied the normal and superconductive state characteristics (resistivity, Hall coefficient, heat capacity and magnetization) of model strongly correlated electronic systems $Lu_xZr_{1-x}B_{12}$ with cooperative Jahn-Teller instability of the boron rigid cage and with dynamic charge stripes. It was found that these metals are *s-wave* dirty limit superconductors with a small mean free path of charge carriers $l$ = 5 - 140 Å and with a Cooper pair size changing non-monotonously in the range 450 - 4000 Å. The parent $ZrB_{12}$ and $LuB_{12}$ borides are type-*I* superconductors, and Zr to Lu substitution induces a type-*I* - type-*II* phase transition providing a variation of the Ginzburg-Landau-Maki parameter in the limits $0.65 \leq \kappa_{1,2} \leq 6$. We argue in favor of the two-band scenario of superconductivity in $Lu_xZr_{1-x}B_{12}$ with gap values $\Delta_1$ ~ 14 K and $\Delta_2$ ~ 6 - 8 K, with pairing corresponding to strong coupling limit ($\lambda_{e-ph}$ ~ 1) in the upper band, and to weak coupling ($\lambda_{e-ph}$ ~ 0.1 - 0.4) in the lower one. A pseudogap $\Delta_{ps-gap}$ ~ 60 - 110 K is observed in $Lu_xZr_{1-x}B_{12}$ above $T_c$. We discuss also the possibility of anisotropic single-band superconductivity with stripe-induced both pair-breaking and anisotropy, and analyze the origin of unique enhanced surface superconductivity detected in these model compounds.

# I. Introduction.

During the last 35 years studies of high temperature superconductivity (HTSC) were developed on cuprates (see, for example, [1]–[7]), Fe-based pnictides and chalcogenides [8]–[17] and MgB$_2$ [18]–[23] to understand the mechanisms at work. In these unconventional and conventional HTSC a lot of unusual phenomena have been found including charge and spin stripes [2]–[4], [7], [12], electron nematic effect [2], [6], pseudogap and strange metal phase [2]–[6], [12]–[14], multiband superconductivity of different type [14]-[24], etc. It was suggested that at least some of these anomalies are closely related with the mechanism underlying the superconductivity enhancement [2]–[6], [12]–[14]. It is believed now that MgB$_2$ clearly belongs to a different class than unconventional HTSCs, being a phonon-driven superconductor, while true, high-temperature superconductivity both in cuprates and Fe-based compounds is a strong correlation phenomenon [2]–[6], [14]. At the same time, however, regardless of that how unique the cuprates may be, these features are not prerequisites for non-phonon high-temperature superconductivity [14]. Thus, it is believed at present that the interplay of different simultaneously active charge, spin, lattice, and orbital interactions plays a key role not only in the formation of a rich variety of phases in phase diagrams of strongly correlated electron systems (SCES), but that they also provide the essential ingredients of HTSC [2]–[6], [12]–[14], [25].

The discovery of superconductivity at $T_c \approx 39$ K in MgB$_2$ [18] stimulated a significant interest in the studies of a wide class of the rare-earth and transition-metal borides. Among them, in the family of high borides RB$_{12}$, zirconium dodecaboride is supposed to be a BCS superconductor with the highest $T_c \approx 6$ K [26], [27]. An important feature established for ZrB$_{12}$ is the key role in formation of Cooper pairs coming from quasi-local vibrations (rattling modes) involving Zr$^{4+}$ ions located within truncated B$_{24}$ octahedrons in the UB$_{12}$-type *fcc* crystal structure [26]–[29] (see Fig. 1). In studies of the Einstein phonon mediated superconductivity in ZrB$_{12}$ the authors of [26]–[33] argue that *s*-wave pairing is characteristic for this compound, and that the Ginzburg–Landau-Maki (GML) parameter κ is located in the nearest vicinity of the threshold value $\kappa_c = 1/\sqrt{2}$. Moreover, a crossover from type-*I* to type-*II/1* superconductivity with temperature lowering was deduced in [27] from heat capacity and magnetization measurements (see inset in Fig. 3b below). In contrast, in [34] the superconductivity in ZrB$_{12}$ was interpreted in terms of *d-wave pairing* and a *two-gap* type-*II* regime was identified. Additionally, a large size *pseudo-gap* ($\Delta_{ps\text{-}gap} \sim 7.3$ meV) has been detected employing high resolution photoemission spectroscopy in ZrB$_{12}$ above $T_c$, and proximity to the quantum fluctuation regime was predicted from the *ab initio* band structure calculations [35]. Thus, a similarity with cuprate HTSC has led to a revival in interest in this low temperature superconductor.

In the case of RB$_{12}$ the replacement of heavy non-magnetic ions from Zr by Lu produces an approximately 15-fold reduction in superconductivity [28], [33], [36], [37] ($T_c \approx 0.4$ K for LuB$_{12}$), and the origin of this $T_c$ suppression is not cleared up to now for these two compounds with similar conduction band and the same crystalline UB$_{12}$-type structure (fig.1a). Indeed, inelastic neutron scattering studies of the phonon spectra in LuB$_{12}$ and ZrB$_{12}$ have detected noticeable, but not dramatic changes in the position of almost dispersion-less quasi-local mode (15 meV and 17.5 meV, correspondingly [29]), which was proposed to be responsible for Cooper pairing. Only a moderate difference in electron density of states (DOS) of these two compounds is caused by filling the wide enough conduction band (~1.6÷2 eV, [28], [38], [39]) when Lu$^{3+}$-ion is changed to Zr$^{4+}$ in the RB$_{12}$ unit cell, resulting to elevation by about 0.3–0.4 eV of the Fermi level $E_F$ for ZrB$_{12}$ in comparison with LuB$_{12}$ [35], [40]. An evidence for the formation of nodes in the superconducting gap of Zr-rich Lu$_x$Zr$_{1-x}$B$_{12}$ dodecaborides was found in [41] and, then, a *s+d-* to- *s*-wave crossover was observed using these *μSR* measurements. The authors of [41] note that the unusual *transition from nodal to gapped superconductivity* is similar to that observed in iron pnictide superconductors.

Recently it was detected that a Jahn-Teller instability of the rigid boron cage develops in LuB$_{12}$ and is accompanied with formation of *dynamic charge stripes* (see Fig.1a-1c) both in the non-magnetic reference compound as well as in rare-earth dodecaborides with magnetic ions [42]–[46]. Besides, an order-disorder phase transition was found in LuB$_{12}$ at $T^* \sim 60$ K [47] and below $T^*$ an infinite cluster in the filamentary structure of stripes appears in the cage-glass state with random displacement of R-ions from central positions in B$_{24}$ cubooctahedra [44] (see Fig. 1d-1e). Thus, taking into account that the essential ingredients of HTSC (stripes, pseudo-gap, $s+d$- wave superconductivity) are observed also in conventional Lu$_x$Zr$_{1-x}$B$_{12}$ superconductors it is promising to study in detail both the normal and superconducting state characteristics of these non-equilibrium dodecaborides to look for commonalities which may be important for HTSC.

Here we probed the evolution of superconducting and normal state parameters for substitutional solid solutions Lu$_x$Zr$_{1-x}$B$_{12}$ employing resistivity, Hall effect, heat capacity and magnetization measurements. It will be shown below that the purest parent ZrB$_{12}$ and LuB$_{12}$ compounds are inhomogeneous type-*I* superconductors with dynamic charge stripes. The non-magnetic Lu impurity substitution then induces a type-*I* – type-*II* transition in Lu$_x$Zr$_{1-x}$B$_{12}$ which is accompanied with $T_c$ lowering and non-monotonous changes in the coherence length $\xi = 450 - 4000$ Å exhibiting a minimum near the Lu percolation threshold $x_c \sim 0.23$. We discuss two alternative scenarios of superconductivity in the dirty limit: (*i*) two-bands with the *s*-wave pairing and (*ii*) single-band *s*-wave anisotropic superconductivity in the presence of strong stripe induced pair-breaking mechanism along <110> direction. A pseudogap state with $\Delta_{ps\text{-}gap} = 60 - 110$ K is detected above $T_c$ in all studied Lu$_x$Zr$_{1-x}$B$_{12}$ dodecaborides.

## II. Experimental details

The high quality single crystals of Lu$_x$Zr$_{1-x}$B$_{12}$ solid solutions were grown by crucible-free inductive floating zone technique in the inert gas atmosphere (see [48] for more detail). The quality and single phase of crystals were controlled by x-ray diffraction. In order to control the composition of samples we used additional optical emission spectral analysis and microanalysis [48]. The obtained values of lattice constant $a(x)$ are shown in fig.1f in combination with the schematic view of the UB$_{12}$–type crystal structure (fig.1a) and the location of dynamic charge stripes detected in LuB$_{12}$ single crystals (fig.1b-1c, [43]). The heat capacity was measured using a Quantum Design PPMS-9 installation in the Shared Facility Centre of P.N. Lebedev Physical Institute of RAS in the temperature range 0.3 -300 K and in magnetic fields up to 9 T. Field and temperature dependences of magnetization were recorded by a Quantum Design MPMS-5. To measure the magnetic characteristics down to very low temperatures (~50 mK) an original *ac*-susceptometer based in a dilution $^3$He -$^4$He mini-refrigerator was applied [49]. For measurements of resistivity and Hall effect we used an original setup described in [50].

## III. Experimental results and data analysis.
### A. Resistivity.

Figures 2a, 2b show the temperature dependencies of resistivity $\rho(T)$ of several studied Lu$_x$Zr$_{1-x}$B$_{12}$ crystals. The $\rho(T)$ curves exhibit a typical metallic behavior with an about linear temperature dependence $\rho = \rho_0 + \alpha T$ in the range $T > 80$ K and a rather small residual resistivity ratio $\rho(300\text{ K})/\rho_0 = 1.8$–20. The large scale plot on Fig. 2c demonstrates the superconducting transitions of Zr-rich Lu$_x$Zr$_{1-x}$B$_{12}$ samples. For Lu$_x$Zr$_{1-x}$B$_{12}$ single crystals with $x \neq 0$ we observed a wide enough resistivity transition with a width $\Delta T_c^{(\rho)} \sim 0.1$–0.4 K as well as a non-monotonous $\rho(T)$ behavior near $T_c$ (Fig. 2c). Both the residual resistivity $\rho_0(x)$ and the slope $\alpha(x)$ change non-monotonously with a maximum near $x \sim 0.5$, and an additional singularity is observed in vicinity of the Lu percolation threshold $x_c \sim 0.23$ (Fig. 3a). A detailed analysis of the resistivity temperature dependencies, similar to that one presented for LuB$_{12}$ in [44], is outside the scope of this paper and will be published elsewhere. Note, that the anomaly at $x_c$ may be attributed to pinning of the dynamic charge stripes on Lu impurities leading to formation of an infinite cluster (filamentary structure of fluctuating electron density) near the percolation threshold.

### B. Magnetoresistance and Hall effect.

Field and temperature dependencies of magnetoresistance (MR) $\Delta\rho/\rho(H,T)$ and Hall coefficient $R_H(H,T)$ have been studied here for a number of $Lu_xZr_{1-x}B_{12}$ crystals (see, for example, Figs. 4a-4c and Figs. S8-S9 in [48]). A quadratic MR behavior $\Delta\rho/\rho \approx \mu_D^2 H^2$ was found and dependencies of the drift mobility of charge carriers $\mu_D(T)$ (Fig. 4d) and $\mu_D(x)$ (Fig. 4e) have been deduced from the data obtained. It is shown in Fig. 4c that in solid solutions $Lu_xZr_{1-x}B_{12}$ the Hall coefficient $R_H(H,T=4.2K)$ is practically field independent parameter and Hall mobility $\mu_H = R_H/\rho$ temperature and concentration changes, $\mu_H(T)$ and $\mu_H(x, T=4.2K)$ have been estimated (see e.g. Figs. 4d and 4e, correspondingly). The Hall factor $\mu_H(x)/\mu_D(x)$ varies in the range 0.45-0.9 and taking the effective mass $m^* \sim 0.7\, m_0$ [51], [52], [53] we estimate also the average relaxation time of charge carriers $\tau(x)$ (fig.4e). It is worth noting that the small enough magnitude of the Hall and drift mobility 10-300 cm$^2$/(V·s) (low magnetic field regime $\omega_c\tau \ll 1$, where $\omega_c$ is the cyclotron frequency) is controlled both by the substitutional disorder and random displacement of heavy Zr/Lu ions from the central positions in the $B_{24}$ cuboctahedrons in $Lu_xZr_{1-x}B_{12}$ solid solutions reaching the minimum values near $x \sim 0.5$ (Fig. 4e).

### C. Specific heat.

The heat capacity temperature dependencies $C(T)$ of several investigated $Lu_xZr_{1-x}B_{12}$ single crystals are shown in Fig. 5a. Panels (b) and (c) in Fig. 5 highlight the zero-field normalized heat capacity behavior in the superconducting state. In addition, Fig. 5a shows also the $C(T)/T = f(T^2)$ curves measured in magnetic field of about 1 kOe in which the superconductivity of studied compounds is completely suppressed. The plot is commonly used to determine the Sommerfeld coefficient $\gamma$ of the electronic heat capacity. As can be seen in Fig. 5a, a gradual diminution of heat capacity at temperatures between 300 K and 50 K is followed by a sharp almost step-like decrease with a typical Einstein-type $C(T)$ dependence below 40 K. It is worth noting that although in the normal state at $T > 30$ K the $C(T)$ curves of all $Lu_xZr_{1-x}B_{12}$ samples are almost identical in the double logarithmic plot used in Fig. 5a, the position of the step-like $C(T)$ anomaly shifts up along the $T$ axis when the concentration $x$ decreases (see e.g. Fig. S10 [48]).

A comparative analysis of contributions to specific heat in the normal state of $Lu_xZr_{1-x}B_{12}$ was carried out in the framework of the approach developed for cage-glass systems in [33], [44], [47]. The Debye temperatures $\theta_D(Lu^NB_{12}) = 1190$ K and $\theta_D(Zr^NB_{12}) = 1500$ K were used to estimate the contribution from the boron sublattice component $C_{ph}$ [28], [33], [44], [47], [54]. After elimination both the electron heat capacity $C_{el} = \gamma T$ and Debye component $C_D$ was the rest $C_{exp}-C_{el}-C_D$ approximated by a sum of Einstein oscillators $C_E$ attributed to Zr and Lu vibrations in oversized $B_{24}$ octahedron cages,

$$\frac{C_E}{T^3} = \frac{3R}{T_E^3}\left(\frac{T_E}{T}\right)^5 \frac{e^{-\frac{T_E}{T}}}{(1-e^{-\frac{T_E}{T}})^2} \tag{1}$$

and Schottky type contributions

$$C_{Schi} = RN_i g_{0i} g_{1i} \left(\frac{\Delta E_i}{k_B T}\right)^2 \frac{e^{\frac{\Delta E_i}{k_B T}}}{(g_{0i} e^{\frac{\Delta E_i}{k_B T}} + g_{1i})^2} \tag{2}$$

($R$- gas constant, $T_E$- Einstein temperature, $g_{0i}$ and $g_{1i}$ are the degeneracies, $i = 1, 2$ and $\Delta E_i$ are the splitting energies) provided by $N_i$ two-level systems (TLS) arranged in double-well potentials (DWP) in vicinity of the randomly distributed Zr/Lu ions (see fig.1e). It was found in [47] that the TLS are created resulting from displacement of these heavy ions from the central positions inside $B_{24}$ cavities in the disordered cage-glass phase of the $RB_{12}$ compounds [47], [55], [56]. The examples of the $Lu_xZr_{1-x}B_{12}$ data analysis are presented in fig. S11 in [48]. A detailed analysis of the specific heat temperature dependences is outside the scope of this article and will be published elsewhere. It is worth noting here that the splitting energy $\Delta E_2$ for $TLS_2$ in Eq.(2), or, in another words, the barrier in the double-well potential in vicinity of heavy ions (fig.1e) should be attributed to the pseudogap $\Delta_{ps\text{-}gap} = \Delta E_2 \cong T^*$, and, hence, when the temperature decreases below $T^*$ the freezing in potential minima of DWP is the main factor which is responsible for the local disorder in the position of Zr/Lu ions. As detected in the $Lu_xZr_{1-x}B_{12}$ family the pseudogap changes $\Delta_{ps\text{-}gap}(x)$ are shown in fig. 6a.

The results of heat capacity measurements at low temperatures and in small magnetic fields which just destroy superconductivity are presented in fig. 7. For comparison, these curves are shown for samples $x = 0.1, 0.46$ and $0.74$ with a significantly different $T_c$ [see panels (a), (b), and (c), respectively] in coordinates $C(T, H_0)/T$ vs $T$ (see also fig. S12 in [48]). Note that the superconducting transition temperature $T_c(x)$ (Fig. 6a) deduced from the heat capacity data are similar to those obtained both from resistivity (Fig. 2c) and field-cooled ($H = 1\text{-}6$ Oe) magnetization curves (see fig. 2d). Apart from $T_c(x)$ changes in $Lu_xZr_{1-x}B_{12}$ there are also differences related both lowering of the jump amplitude $\Delta C$ near $T_c$ and the broadening of this anomaly (see fig. S13 in [48]). For all samples except that one with $x = 0.03$ the linear dependencies in the upper-left part of fig. 5a allow to estimate the $\gamma(x)$ values (see fig. 8a), whereas the low temperature specific heat of the $x = 0.03$ sample is obviously influenced by a moderate additional magnetic contribution.

It should be mentioned that magnetic and non-magnetic $Lu_xZr_{1-x}B_{12}$ samples have been found previously [37], [41], [57] depending on both the crystal growth conditions and the location of Lu-ions in crystals. Moreover, the formation of magnetic nanodomains near the pairs of randomly distributed Lu-ions in the $Lu_xZr_{1-x}B_{12}$ solid solutions was concluded to be responsible for the low temperature magnetic component of the heat capacity [41], [57]. It is well-known at present that any magnetic defects, clusters, and spin glass behavior can result into a specific heat enhancement [58] and lead in some cases to a false indication of heavy fermion behavior [59], [60]. In such cases a detailed investigation of magnetic field changes of the low temperature heat capacity can help to identify the nature of the enhancement. For this reason we have carried out field dependent heat capacity measurements of the magnetic crystal $x = 0.03$ to separate the electronic and magnetic contributions. The obtained magnetic component which shifts up along the temperature axis when the magnetic field increases, prevails essentially the electronic Sommerfeld term ($\gamma(x=0.03) \sim 4.3$ mJ/(mole·K$^2$)) and demonstrates a moderate increase in external magnetic field (see fig. S14 in [48] for more details).

The specific heat results obtained in the normal and superconducting states (Figs. 5, 7 and fig. S12 in [48]) were used to determine the thermodynamic critical field $H_{cm}(T)$ within the framework of standard relations

$$-1/2\ \mu_0 V H_{cm}^2(T) = \Delta F(T) = \Delta U(T) - T\Delta S(T) \quad (3)$$

$$\Delta U(T) = \int [C_s(T') - C_n(T')]dT' \quad (4)$$

$$\Delta S(T) = \int dT'[C_s(T') - C_n(T')]/T' \quad (5),$$

where $F$ and $U$ denote the free and internal energies, $S$ the entropy, $V$ the molar volume, and the indices $n$ and $s$ correspond to characteristics of the normal and superconducting phases of $Lu_xZr_{1-x}B_{12}$. The integration was carried out in the temperature range from $T$ to $T_c$. Before integration the specific heat data in the normal and superconducting states were approximated by polynomials of the 4$^{th}$ order. Fig. 9 shows the dependencies of the thermodynamic $H_{cm}(T)$ and upper $H_{c2}(T)$ critical fields, respectively, resulting from the heat capacity analysis of studied crystals. Fig. 10a presents both the $H_{cm}(0)$ values obtained by extrapolation of $H_{cm}(T)$ curves in the framework of the standard Bardeen-Cooper-Schriffer (BCS) relation [61]

$$H_{cm}(T)/H_{cm}(0) = 1.7367(1-T/T_c)[1-0.327(1-T/T_c)-0.0949(1-T/T_c)^2] \quad (6)$$

and $H_{c2}(0)$ magnitudes defined within the framework of formula used in [62]

$$H_{c2}(0) = -0.69 T_c (dH_{c2}/dT)_{T=T_c} \quad (7)$$

with derivatives $dH_{c2}/dT$ at $T = T_c$ obtained from the experimental data. Using the electronic specific heat coefficient $\gamma(x)$ (Fig. 8a), the density of electronic states (DOS) at the Fermi level $N_b(E_F) = 0.122$ (eV·atom)$^{-1}$ for ZrB$_{12}$ and $N_b(E_F) = 0.108$ (eV·atom)$^{-1}$ for LuB$_{12}$ known from band structure calculations [35], [40], [53], [63]–[66] and the relation

$$\gamma = 1/3\pi^2 k_B^2 N_b(E_F)(1+\lambda_{e\text{-}ph}) \quad (8)$$

($k_B$ - Boltzmann constant), we have estimated the electron-phonon interaction constant $\lambda_{e\text{-}ph}(x)$ (see inset in fig. 8b, $N_b(E_F)$ values for various Lu content were taken by the linear interpolation between DOS of LuB$_{12}$ and ZrB$_{12}$) which for parent compounds LuB$_{12}$ and ZrB$_{12}$ is in good agreement with results of [26]–[28]. For independent evaluation of the renormalized $N(E_F) = N_b(E_F)(1+\lambda_{e\text{-}ph})$ and $\lambda_{e\text{-}ph}$ we use also the relation

$$\Delta C/T_c = 0.95\pi^2 N(E_F) \quad (9),$$

which links the jump of heat capacity at $T_c$ with $N(E_F)$ in the family Lu$_x$Zr$_{1-x}$B$_{12}$ superconductors (see $\Delta C/T_c(x)$ dependence in fig. 8b). Then, from BCS relations

$$\Delta(0) = (2\pi N(E_F))^{-1/2} H_{cm}(0) \quad (10)$$

$$\xi(0) = (\Phi_0/2\pi H_{c2})^{1/2} \quad (11)$$

$$\kappa_1(T) = 2^{-1/2} H_{c2}(T)/H_{cm}(T) \quad (12),$$

where $\Phi_0$ denotes the flux quantum, the GLM parameter $\kappa_1(T)$ [67] (see fig. S16 in [48]), the single-band average superconducting gap $\Delta(0)$ (Fig. 6b), the coherence length $\xi(0)$ and the penetration depth $\lambda(0) = \kappa_{1,2}(0)\cdot\xi(0)$ (Fig. 10b) could be calculated. For a few Lu$_x$Zr$_{1-x}$B$_{12}$ magnetic crystals the evaluation of the Sommerfeld coefficient $\gamma$ was obtained from the dimensionless ratio $\gamma T_c^2/\mu_0 V H_{cm}^2(0) = const$ [26], [68], [69]. This last invariant was estimated to be 1.95±0.15 for ZrB$_{12}$ and ranging in the interval 2-2.2 for other Lu$_x$Zr$_{1-x}$B$_{12}$ compositions in accordance with s-wave pairing [68] and being only slightly above the BCS value 0.17 [70]. An average gap ratio of $2\Delta/k_B T_c = 3.7\pm0.15$ was found for all Lu$_x$Zr$_{1-x}$B$_{12}$ samples except LuB$_{12}$ where $2\Delta/k_B T_c = 3.2\pm0.1$ was calculated. Note that in the case of ZrB$_{12}$ and LuB$_{12}$ the gap to $T_c$ ratio coincides very well with the results obtained from the heat capacity analysis of [26]–[28], [36] and for ZrB$_{12}$ it exceeds slightly the value of 3.52 of the BCS model. It is also worth noting that surface sensitive techniques provide much more higher ratios, 4.8 in Andreev reflection experiments [32], 4.15 [71] and 4.75±0.1 [30] in the point-contact and tunnel spectra measurements, correspondingly, and 4.77±0.04 in ZrB$_{12}$ powder magnetization studies [72]. These differences have been pointed out by Tsindlekht et al. [30], [31], [73] and explained by enhanced surface characteristics of ZrB$_{12}$ leading to rather different superconducting properties of bulk [26], [27], [30], [33], [65], [74] and surface [30]–[32], [34], [71]–[73].

When discussing finally the superconducting characteristics in terms of *s*-wave weak coupling BCS superconductors ($\lambda_{e-ph} \leq 0.4$, see inset in fig. 8b) note that even in case of ZrB$_{12}$ the quality of the *α*-model fit

$$C_s(T)=A_0 T^{-3/2} exp(-\Delta(0)/k_B T) \quad (13),$$

is not adequate to describe the electronic specific heat, and the two-gap or anisotropic gap scenarios (see [24], for details) look like a better fitting approximation (see figs. 5b-5c). Fig. 6b shows the changes of the two-gap *α*-model parameters $\Delta_1(0)$ and $\Delta_2(0)$ in the family Lu$_x$Zr$_{1-x}$B$_{12}$ and the inset presents the *x*-dependence of the relative weight $n_2(x)$ of the small-gap component.

### D. Magnetization.

Below the transition temperature $T_c$ a diamagnetic response is detected on magnetization curves $M(T)$ (fig. 2d) in very small magnetic fields (1-6 Oe). Note, that a strong paramagnetic Meissner effect (PME) has been observed previously in studies of ZrB$_{12}$ single crystals [75], and the PME was related to superconducting surface states [76]. However, within the limit of experimental accuracy we have not found any influence of PME on the magnetic characteristics of the superconducting state of studied Lu$_x$Zr$_{1-x}$B$_{12}$ samples. An increase of external magnetic field up to 1 kOe leads to the appearance of features on $M(H, T_0)$ curves which are typical for type-*II* superconductors. Indeed, a linear rise of the diamagnetic magnetization is observed in the range below the lower critical field $H < H_{c1}$ corresponding to Meissner phase with about total (~100%, see also fig. S15 in [48]) response, and above $H_{c1}$, in the mixed state, $M(H)$ decreases dramatically until the transition to normal state at the upper critical field $H_{c2}$ occurs. Fig. 11 demonstrates the diamagnetic $M(H, T_0)$ dependences as obtained for Lu$_x$Zr$_{1-x}$B$_{12}$ samples with $x$ = 0.04, 0.1 and 0.2 (panels a, b and c, respectively). The procedure usually applied for the extraction of critical fields is shown in the insets of fig. 11, where the intersection points of linear asymptotics marked as $H_{c1}$ and $H_{c2}$ are shown for various temperatures. The values of $H_{c1}$ were corrected to the demagnetization factor. The received behavior of the critical fields for Zr-rich crystals of Lu$_x$Zr$_{1-x}$B$_{12}$ is presented in fig. S17 in [48]. The normalized dependences $h_{c1} = H_{c1}/H_{c1}(0)$ vs. $(T/T_c)$ and $h_{c2} = H_{c2}/H_{c2}(0)$ vs. $(T/T_c)$ scaled very well for all Zr-rich samples and these almost coincide with each other (fig. S17 in [48]) and with commonly used phenomenological and BCS approximations.

The analysis of magnetization was carried out based on formulas which are well-known from the Abrikosov theory of type-*II* superconductivity [77]

$$-4\pi M = (H_{c2} - H)/((2\kappa_2^2 - 1)\beta_\Delta) \quad (14)$$

$$H_{c1}(T) = H_{c2}/2\kappa_2^2 (\ln\kappa_2 + a) \quad (15),$$

where $\kappa_2$ is the GLM parameter [67], [77], [78], $\beta_\Delta = 1.16$ the coefficient corresponding to a triangular lattice of Abrikosov vortices, and $a$ the constant depending on impurity concentration. Presented in fig. 11 are the linear dependences of magnetization $M(H)$ in the superconducting phase near $H_{c2}$ which allow to derive the $\kappa_2(T)$ behavior within the framework of Eq.(14) (see fig. S16 in [48]). Then, the extrapolation to zero temperature provides values of $\kappa_2(0)$ (fig. 3c) and $a$ (not shown) parameters. In addition, we use Eq.(11) to estimate the coherence length $\xi(0)$ and the penetration depth $\lambda(0) = \kappa_2(0) \cdot \xi(0)$; $H_{c1}(0)$ (see fig. 10b) was defined from magnetization data of fig. 11. The comparison of GLM parameters $\kappa_1(T)$ and $\kappa_2(T)$ [67], [77], [78] for Lu$_x$Zr$_{1-x}$B$_{12}$ crystals with $x$ = 0.104 and 0.17 obtained from the analysis of heat capacity (Eq. (12)) and magnetization (Eq. (14)), respectively, shows that $\kappa_1$ and $\kappa_2$ differ noticeably (up to 20%, see fig.

S16 in [48]), arguing in favor of inhomogeneous superconductivity in these non-equilibrium compounds with dynamic charge stripes. According with the recommendations of [70] the heat capacity data were considered as superior to the results of magnetization measurements in determining of $H_{c2}(T)$. Taking into account strong enhancement of superconductivity in the surface layer of ZrB$_{12}$ [30]–[32], [34], [71]–[73] we have not studied here the resistivity changes in magnetic field near $T_c$ in the Lu$_x$Zr$_{1-x}$B$_{12}$ crystals.

## IV. Discussion.

### A. Order-disorder transition and pseudogap state.

The formation of two-level systems in dodecaborides has for the first time been observed experimentally in LuB$_{12}$ [47] and ZrB$_{12}$ [33] where Lu/Zr ions are embedded in large size cavities arranged by B$_{24}$ cubooctahedra (see fig.1). In Raman spectra of Lu$^N$B$_{12}$ [47] and Zr$^N$B$_{12}$ [79] with a different isotopic composition of boron (N = 10, 11, nat; nat corresponds to the natural content of boron isotopes: 18.83% $^{10}$B and 81.17% $^{11}$B) it was shown that the Raman response exhibits a boson peak at liquid nitrogen temperatures, and such a feature in the low-frequency range was discussed as a fingerprint of systems with strong structural disorder. To explain the properties of LuB$_{12}$ authors of [47] have proposed a model of cage-glass formation with a phase transition at $T^* \sim 50 - 70$ K, and it was found that the barrier height of the double-well potential $\Delta E_2$ (Fig. 1e) is practically equal to the cage-glass transition temperature $T^*$. In ZrB$_{12}$ the order-disorder transition was found at $T^* \sim 90$-$100$ K [37] and $\Delta E_2$ and $T^*$ parameters are about equal to the pseudogap $\Delta_{ps\text{-}gap} = 7.3\pm2$ meV detected by the high resolution photoemission spectroscopy of zirconium dodecaboride [35]. It was shown in [33], [47] that the temperature lowering at $T < T^*$ leads to displacements of metallic ions from their central positions inside the B$_{24}$ cubooctahedra (see fig. 1d-1e and also [55], [56]). The result is a static disorder in the arrangement of Lu$^{3+}$(Zr$^{4+}$) ions while maintaining the rigid covalent boron framework (the so-called cage-glass state).

The presence of two-level systems with a barrier of $\Delta E_2 \sim 90$ K was reliably demonstrated also in cage-glass compound LaB$_6$ and in Ce$_x$La$_{1-x}$B$_6$ solid solutions basing on low-temperature heat capacity measurements [80], [81]. Furthermore, a pseudogap [82] and a low-frequency peak in inelastic light scattering spectra [83], [84] were found in LaB$_6$ and the barrier in the DWP was attributed to the formation of a pseudogap in disordered metallic systems [80], [81]. Later on similar conclusions were made for YB$_6$ where the cage-glass transition at $T^* \sim 50$K was found to be accompanied with the appearance of a DWP barrier of $\Delta E_2 \sim 50$ K and it was discussed in terms of pseudogap emergence [85]. It is worth noting that all these high borides are characterized by small enough residual resistivity values located in the range 0.01-50 µOhm·cm and their conduction band width is about 1.6-2 eV (see e.g. [28], [38], [39], [65], [86]), so the pseudogap may be considered as a merit of disorder and anharmonicity in these good metals.

The $\Delta_{ps\text{-}gap}(x) = \Delta E_2(x)$ changes in the Lu$_x$Zr$_{1-x}$B$_{12}$ family are detected from the analysis of heat capacity (see, for example, fig. S10 in [48]) elucidating the location of the pseudogap area just above the superconducting state on the phase diagram of these non-equilibrium metals (fig. 6). A source of lattice instability in LuB$_{12}$ and another RE and transition metal dodecaborides is related to the cooperative Jahn-Teller instability of B$_{12}$ clusters (ferro-distortive effect), which manifests both the emergence of small static distortions of the *fcc* lattice and the formation of dynamic charge stripes along unique [110] direction depending from the distribution of impurities and imperfections [42]–[45], [87], [88]. According to the conclusions [43], [44], [87] the infinite cluster of stripes appears only in the disordered phase below $T^*$, but the *fcc* lattice instability develops in RB$_{12}$ with temperature lowering well above $T^*$ approaching the Ioffe-Regel limit near $T_E \sim 150$K, where the vibrational density of states reaches the maximum [47] and strong changes appear both in structural parameters and characteristics of the atomic dynamics [44], [88]–[90]. It is worth noting finally in this section that the pseudogap $\Delta_{ps\text{-}gap}(x)$ decreases only slightly in the range $0 \leq x \leq 0.9$ in $Lu_xZr_{1-x}B_{12}$ where we observe an order of magnitude $T_c$ changes, but for $LuB_{12}$ crystals $\Delta_{ps\text{-}gap}$ depresses essentially taking values of 60-80 K depending from the concentration of boron vacancies (see fig. 6a and [47] for more details).

### B. Lengths and limits.

When discussing the superconducting characteristics of the parent $ZrB_{12}$ compound it should be pointed out that since zirconium dodecaboride has a GLM parameter $\kappa$ very close to the threshold value $\kappa_c=1/\sqrt{2}$, small changes in the sample preparation process (e.g., defect concentration and thermal treatment) result in a variation of $\kappa$ between type-*I* and type-*II* superconductivity (see fig. 3b). In our $ZrB_{12}$ samples typical $\kappa$ values were detected between 0.8 and 1.14 [74], but the purest crystals are type-*I* superconductors with $\kappa \sim 0.65$ [26], [27]. Note, that the location of $ZrB_{12}$ near $\kappa_c$ allows detecting both type-*I* - type-*II/1* [26], [27] and type-*II* - type-*II/1* [75], [91] crossovers with the temperature lowering (see inset in fig. 3b and [26], [27], [91] for more details). Moreover, the growth of $ZrB_{12}$ crystals is a very complicated process [48], so, the coexistence of type-*I* and type-*II* superconducting areas are typical, and this kind of inhomogeneity of crystals was directly observed in the *μSR*-studies of zirconium dodecaboride [92]. The correlation length and the penetration depth estimated in the present study for $ZrB_{12}$ are in the range $\xi(0) \sim \lambda(0) \sim 680\text{-}780$ Å (Fig. 10b). Our evaluation of the mean free path of charge carriers $l(0)$ from residual resistivity $\rho_0$ (Fig. 2), from the Hall coefficient $R_H$ (Fig. 4 and [48]), and from parameters $\xi(0)$ (Fig. 10b) and $\Delta(0)$ (Fig. 6b) lead within the framework of standard relations

$$l = R_H\, m^* v_F/(e\rho_0) \tag{16}$$

$$\xi(0) = \hbar v_F/(\pi\Delta(0)) \tag{17}$$

($v_F$ is the average Fermi velocity (see fig. 8b) and $m^*$ is the effective mass, $m^* \sim 0.7m_0$ [52]) to a value of $l \sim 130$ Å for the studied $ZrB_{12}$ crystals (see fig. 10b). This results to a ratio $\xi/l \sim 5$ validating the "dirty limit" superconductivity even for the "pure" parent compound. Note, that as the obtained Fermi velocity $v_F(ZrB_{12}) \sim 3.5 \cdot 10^7$ cm/sec (Fig. 8b) differs strongly from the rough Drude-type estimation $v_F(ZrB_{12}) \sim 1.9 \cdot 10^8$ cm/sec [34]. Moreover, the average Fermi velocity decreases dramatically with the change of lutetium concentration in the range $x(Lu) < 0.8$, starting to increase slightly in Lu-rich solid solutions (Fig.8b). On the contrary, the charge transport relaxation time $\tau(ZrB_{12}) = 3.7 \cdot 10^{-14}$ sec derived here (Fig. 4e) is in agreement with the results deduced from optical studies ($\tau = 4.2 \cdot 10^{-14}$ sec, [28], [65]), high resolution photoemission spectroscopy ($\tau = 6 \cdot 10^{-14}$ sec, [35]), and de Haas -van Alphen quantum oscillations measurements ($\tau = 1.7\text{-}7.4 \cdot 10^{-14}$ sec, [52]). It is worth noting also that a huge value of the Fermi surface averaged Dingle temperature $T_D = \hbar/2\pi k_B\tau \sim 70$ K [52] argues in favor of very strong charge carrier scattering in this inhomogeneous superconductor.

It is seen from fig. 3b that in the range $0 < x < 0.95$ the $Lu_xZr_{1-x}B_{12}$ solid solutions are type-*II* superconductors, and the GLM parameter reaches its maximum value $\kappa \sim 6$ in crystals with $x \sim 0.5$ corresponding to the highest substitutional disorder (fig. 3b). Note, that the $\kappa_{1,2}(x)$ dependence (fig. 3b) correlates very well with the residual resistivity $\rho_0(x)$ changes (fig. 3a) and with the relaxation rate $\tau^{-1}(x)$ (Fig. 4e), so, the well-known relation

$$\kappa_d = \kappa_p + 7.53 \times 10^3 \rho_0 \gamma^{1/2} \tag{18}$$

(see, e.g., [93], $\kappa_d$ and $\kappa_p$ are GLM parameters in the dirty and pure limits, correspondingly) is at least valid qualitatively. Both in $LuB_{12}$ and in Lu-rich crystals with $x > 0.95$ the type-*I* superconductivity recovers. Indeed, it is certainly demonstrated in fig. 12, that the BCS relation Eq.(6) describes very well the temperature dependence of the critical field only for Lu-rich samples, but the type-*II* behavior $h_{c2}(t) = (1-t^2)/(1+t^2)$, where $h_{c2} = H_{c2}/H_{c2}(0)$ and $t = T/T_c$, becomes valid in the concentration range $x \leq 0.74$. So, Lu-rich compounds are type-*I* superconductors characterized by very small $T_c$ (Fig. 6a) and critical field $H_{cm}$ (Fig. 10b), and with a large enough size of Cooper pairs $\xi(LuB_{12}) \sim 4000$ Å (Fig. 10a). It is worth noting that $\xi/l \sim 36$ is detected for $LuB_{12}$ and this ratio increases strongly reaching the largest value $\xi/l \sim 150$ near $x \sim 0.5$ where $l \sim r(R-R) \sim 5$ Å ($r(R-R)$ is the distance between heavy Zr/Lu ions in the *fcc* lattice, see fig. 1a) and validating the "dirty limit" superconductivity for all $Zr_{1-x}Lu_xB_{12}$ compounds. Evidently these huge $\xi/l$ ratios should be considered as a consequence of the JT lattice instability and of the dynamic charge stripes, resulting to very strong charge carrier scattering in inhomogeneous superconductors with nanoscale electron phase separation (see figs. 1a-1c).

When the Lu content increases in the range $x \leq x_c$ in $Lu_xZr_{1-x}B_{12}$ the $T_c(x)$ (fig. 6a) and correlation length $\xi(x)$ (fig. 10b) demonstrate a moderate decrease which is accompanied by a similar decrease of (*i*) the Sommerfeld coefficient $\gamma(x)$, (*ii*) the concentration of charge carriers $n_e/n_R(x) = 1/(R_H e\, n_R)$ ($n_R$ is the concentration of R-ions in $RB_{12}$) and (*iii*) the jump of heat capacity $\Delta C/T_c(x)$ (Figs. 8a,8b). Taking into account that all these parameters $\gamma(x)$, $n_e/n_R(x)$ and $\Delta C/T_c(x)$ are the characteristics of the renormalized DOS $N(E_F) = N_b(E_F)(1+\lambda_{e-ph})$ it is natural to attribute the weakening of superconductivity to the moderate lowering of $\lambda_{e-ph}(x)$ in the range $x \leq x_c$ (see inset in fig. 8b). We need to note that the conservation of parameter $2\Delta/k_BT_c = 3.7\pm0.15$, which is close to BCS value 3.52, for all $Lu_xZr_{1-x}B_{12}$ with $x \leq 0.95$, from one side, and the good quality scaling detected for the critical fields $h_{c1}(t)$ and $h_{c2}(t)$ in the range $x \leq x_c$ (see fig. S17 in [48]), from another, argue in favor of common BSC type description of dirty limit superconductivity in Zr-rich solid solutions.

### C. Scenarios of superconductivity.

Properties of the superconducting state of conventional, single-band, electron-phonon superconductors are well-described within isotropic Eliashberg theory with a single electron-phonon spectral density $\alpha^2(\omega)F(\omega)$ for the average interaction over the Fermi surface [69], [94]. Despite the Eliashberg function for various electrons on the Fermi surface is anisotropic leading to energy gap anisotropy, usually in the dirty limit superconductors where the electron mean free path is much smaller than the coherence length, only Fermi surface averaging of the electron-phonon spectral density can be used. This is not the case of $Lu_xZr_{1-x}B_{12}$ although the $\xi/l$ ratio ranges in limits 5-150 (fig. 10). Indeed, both of BCS and strong coupling single-band scenarios with the deduced constant $\lambda_{e-ph} \leq 0.4$ (see inset in fig. 8b) and with the pre-exponent $\langle \hbar\omega_{ln}\rangle/k_B \approx \theta_E = 160\text{-}200$ K corresponding to Cooper pairing mediated by quasi-local vibrations (rattling modes) of Zr/Lu ions cannot explain both of the $T_c(ZrB_{12}) \approx 6$ K and the $T_c$ changes in solid solutions. For instance, within the framework of the Allen-Dynes relation for superconducting transition temperature [95]

$$k_B T_C = \frac{\hbar \omega_{\ln}}{1.2} \exp\left[-\frac{1.04(1+\lambda_{e-ph})}{\lambda_{e-ph} - \mu^*(1+0.62\lambda_{e-ph})}\right] \qquad (19)$$

in the case of $ZrB_{12}$ we obtain curious,- small and negative Coulomb pseudopotential $\mu^* \sim -0.02$. On the contrary, when taking the traditional value $\mu^* \sim 0.1$ we obtain $\lambda_{e-ph}(ZrB_{12}) \approx 0.65$, which is similar to the coupling constant calculated in [32], but is not in accord with the experimental heat capacity results (see inset in fig. 8b and [26]–[28]). Similar estimation $\lambda_{e-ph} \approx 0.68$ was found from the analysis of the low temperature optical reflectivity spectra of $ZrB_{12}$ using the isotropic transport Eliashberg function $\alpha^2(\omega)F(\omega)$ which has two peaks around 12 meV and 60 meV [65]. According to [28], [36], [65], [79] a low frequency peak in the Eliashberg function was observed in ZrB$_{12}$ near $\omega_0 \sim 100$ cm$^{-1}$ (~12 meV) and it is located well below the Einstein mode at $\omega_E \sim 140$ cm$^{-1}$ (~17.5 meV) [29]. The value $\lambda_{e\text{-}ph} \approx 0.58$ was deduced from the detailed analysis of heat capacity made in [33] where three Einstein modes $\hbar\omega_{E1}/k_B \approx 200$ K and $\hbar\omega_{E2,3}/k_B \approx 450$ K were found to mediate the electron-phonon interaction in ZrB$_{12}$ resulting to pre-exponent $<\hbar\omega_{ln}>/k_B \approx 368$ K. Note again, that the isotropic single-band models do not discuss the difference between the estimated values $\lambda_{e\text{-}ph} \approx 0.58$-0.68 and the much smaller constant $\lambda_{e\text{-}ph} \sim 0.4$ detected from the Sommerfeld coefficient (Fig. 8 and [26]–[28]).

The situation becomes much more appealing in the two-band scenario which is supported obviously by the two-gap $\alpha$-model fit of the heat capacity in the superconducting state of Lu$_x$Zr$_{1-x}$B$_{12}$ (see, for example, figs. 5b-5c). In this case we found $\Delta_1(0) \sim 14$ K and $\Delta_2(0) \sim 6$ K in Zr-rich crystals with $x < x_c$ detecting a larger gap and a strong coupling limit ratio $2\Delta_1/k_BT_c \approx 4.8$. Similar gap values $\Delta_1(0) \sim 14$ K and $\Delta_2(0) \sim 8$ K were deduced in the *rf*-measurements of ZrB$_{12}$ [96]. The ratio for the smaller gap was estimated to be only $2\Delta_2/k_BT_c \sim 2$, and we suppose an essential interband coupling and impurity scattering in these compounds with structural (cooperative JT- effect of B$_{12}$ clusters [43]–[45], [87]) and electron (dynamic charge stripes, fig. 1a-1c) instabilities. The strong difference between the larger and smaller gap ratio is similar to that one observed previously in archetypal two-band superconductor MgB$_2$, in Lu$_2$Fe$_3$Si$_5$, etc. (see, for example, [24] for review). Moreover, the larger gap $\Delta_1(0) \sim 14$ K is about equal to values ~14-15K observed for ZrB$_{12}$ in surface sensitive experiments [30], [32], [71], [72]. The unique enhanced surface superconductivity in ZrB$_{12}$ [31], [73] was explained in [97] in framework of the self-organized percolative theory of superconductivity [98] suggesting the formation of a filamentary network and disorder-enhanced electron-phonon coupling at the surface. The authors [72] considered the possibility of non-adiabatic coupling of charge carriers to the crystal lattice appearing in ZrB$_{12}$ close to the surface. The difference between the surface and bulk state in ZrB$_{12}$ has been studied in [99] using high resolution x-ray spectroscopy which reveals boron deficiency at the surface while the bulk is stochiometric. Correspondingly, in the two-band scenario of superconductivity in ZrB$_{12}$ one needs to suppose that the enhanced surface characteristics may be related to the suppression of the smaller superconducting gap in the surface layer.

It is interesting to note that similar values $2\Delta_1/k_BT_c \approx 4.8$ and $2\Delta_2/k_BT_c \sim 2$ were predicted for the gap ratios in two-band *s*-wave superconductors in detailed Eliashberg calculations developed in [94]. For the coupling constants $\lambda_{11} = 1$ and $\lambda_{22} = 0.1$ in the upper and lower bands authors [94] obtained $\lambda_{12} \approx 0.4$ and $\lambda_{21} \approx 0.3$ for the off-diagonal interband elements of the electron-phonon interaction (see fig. 3 in [94]) confirming the case of an essential interband coupling in Lu$_x$Zr$_{1-x}$B$_{12}$. A comparison of the calculated and experimental cyclotron mass made in studies of the de Haas-van Alphen effect showed unusually large electron-phonon interaction on the neck ($\lambda_{e\text{-}ph} \approx 0.95$) and box ($\lambda_{e\text{-}ph} \approx 1.07$) sections of the Fermi surface of ZrB$_{12}$ [52]. Similar estimation $\lambda_{e\text{-}ph} \approx 1.0 \pm 0.3$ was obtained in optical measurements [65] indicating a strong coupling in ZrB$_{12}$. Note also that in the two-band scenario the volume averaged quasiparticle density of states at the Fermi surface is expected to be not large, providing small enough values of the Sommerfeld coefficient, but the DOS in the upper band at temperatures above $T_c$ should be much higher. So, the contradiction between small experimental $\gamma$ and the deduced $\lambda_{e\text{-}ph} \sim 0.4$ values, from one side, and large DOS and upper band coupling $\lambda_{e\text{-}ph} \approx 1.0$, from the other seems to be removed. Taking into account that the zeroing of heat capacity in the superconducting state is observed up to threshold-temperature ~$0.2T_c$ (figs. 5b-5c) we conclude also in favor of the *s*-wave superconductivity, while in *d*-wave systems the gap nodes allow a significant occupation of the excitation spectrum at any finite temperature, which makes $C/\gamma T$ to increase strongly even at very low temperatures [24]. Another argument in favor of *s*-wave superconductivity in Lu$_x$Zr$_{1-x}$B$_{12}$ is the small enough dimensionless ratio $\gamma T_c^2/\mu_0 V H_{cm}^2(0) = 1.95$-$2.2$ which is nearly twice smaller as that predicted for $d$-wave scenario in [68].

The two-band scenario of superconductivity in $Lu_xZr_{1-x}B_{12}$ has been proposed on the basis of $\mu SR$ experiments to extract the superfluid density [41]. The best fits to the $\mu SR$ data on samples with low values of $x$ were obtained with two-band models, with an $s+d$-model giving the best fit [41]. However, there are many possible types of multi-gap behaviour that may be realised outside the context of these simple models, and in particular the inclusion of impurity scattering and interband coupling can affect the detailed temperature dependence of the superfluid stiffness [24,94,100]. Therefore the $\mu SR$ data [41], though strongly pointing to two-band superconductivity, could very well be consistent with two-band $s$-wave superconductivity in the presence of strong impurity scattering and interband coupling.

It is worth noting that in two-gap superconductors the specific effect of anisotropy may be attributed not only to the anisotropic electron-phonon interaction spectral function, but also could be described in terms of the anisotropy of Fermi velocity [101]. For instance, the upper critical field $H_{c2}(T)$ depends on the orientation of applied magnetic field $H$ due to the anisotropy of Fermi velocity and its positive curvature near $T_c$ may be considered as a measure of $v_F$ anisotropy [24]. Fig. 12 shows that the positive curvature emerges and grows with decreasing of Lu content in $Lu_xZr_{1-x}B_{12}$ solid solutions. The positive curvature of $H_{c2}(T)$ near $T_c$ is conserved in the range $x < 0.5$ (see fig. S17 in [48]) giving evidence in favor of anisotropic Fermi velocity. In our opinion, a huge (about 5 times) decrease of average Fermi velocity $v_F(x)$ (Fig. 8b) also argues in favor of strong interband coupling, impurity scattering and very strong anisotropy of superconductivity in Zr-rich $Lu_xZr_{1-x}B_{12}$. Note, that the upper critical field $H_{c2}$ was one of the first properties of $MgB_2$, for which unconventional behavior was discerned, namely by a pronounced positive curvature of its temperature dependence near $T_c$ [102]–[104]. Since then, this effect has been discovered in many compounds and considered as a confirmation of two-band superconductivity.

Following to [24] it should be pointed out, however, that the same effect on $H_{c2}(T)$ occurs in *anisotropic single-band superconductors*. Moreover, different energy gap values, from which the smaller ones dominate the low- and the larger ones the high temperature behavior, are also found in anisotropic single-band superconductors. Accordingly, the resulting curves could easily resemble the two-band behavior [24]. Though it seems difficult to imitate the extreme two-band case, in which the contributions of the two gaps can still be distinguished (i.e. when interband effects are very weak). Most experimentally observed curves showing almost a linear or slightly convex behavior over large parts of the temperature range fit the anisotropic single-band model well, as was e.g. demonstrated for $MgB_2$ [105] (see also [24] for review). So, the anisotropic single-band scenario can not be ruled out for $Lu_xZr_{1-x}B_{12}$ and it should be verified in future in detailed magnetic field measurements of the heat capacity, superfluid density, etc.

Concluding the section we need to discuss the physical meaning of the critical behavior near $x_c \sim 0.23$ where anomalies of residual resistivity $\rho_0(x)$ and $\alpha(x)$, $B_0(x)$ and $T_0(x)$ (Fig. 3), of the Hall mobility $\mu_H(x)$ and relaxation time $\tau(x)$ (Fig. 4), $T_c(x)$, energy gaps $\Delta_{1,2}(x)$ and the relative weights of superconducting components $n_i(x)$ (Fig. 6), of upper critical field $H_{c2}(x)$ and correlation length $\xi(x)$ (Fig. 10) are observed. Taking into account that fast fluctuations of electron density (dynamic charge stripes detected near the RE ions in the *fcc* lattice of both $LuB_{12}$ [42]–[44], [87] and another RE dodecaborides [45], [46]) and that an infinite cluster appears in the disordered cage-glass phase below $T^* \sim 60$ K it seems natural to associate the anisotropy of superconductivity with these inhomogeneities and nanoscale phase separation in the matrix of $RB_{12}$. Moreover, in the filamentary structure the electron density fluctuations with frequency ~240 GHz (~1 meV) [45] are arranged in chains emerging near the RE ions which form channels along <110> directions [43]–[46] and the Cooper pair breaking effect of these dynamic stripes becomes decisive. At low temperatures $T < T^*$ the infinite cluster develops and enforces above the percolation threshold ($x \geq x_c$ ~0.23), enhancing the pair breaking and depressing the superconductivity. Thus, even in the case of single-band superconductivity the anisotropy which is not removed in the dirty limit system with $\xi/l$ = 5-150 may be also explained by the cooperative JT instability of the boron sub-lattice which is accompanied with formation of a filamentary structure of fluctuating electron density in these inhomogeneous $Lu_xZr_{1-x}B_{12}$ solid solutions.

## V. Conclusions

Normal and superconducting state characteristics have been studied in model strongly correlated electronic systems $Lu_xZr_{1-x}B_{12}$ with cooperative Jahn-Teller instability of boron sub-lattice and with nanoscale electron phase separation in form of dynamic charge stripes. It was shown that the purest $ZrB_{12}$ and $LuB_{12}$ crystals are type-*I* superconductors, and that Zr to Lu substitution induces immediately the type-*I* –type-*II* transition whereas GLM parameters $\kappa_{1,2}$ change in the range $\kappa_{1,2} \leq 6$. It was found that $Lu_xZr_{1-x}B_{12}$ are dirty limit superconductors with short enough mean free path $l$ = 5-140 Å and a coherence length changing non-monotonously in the range 450-4000 Å (the ratio $\xi/l$ = 5-150). The most likely scenario proposed for these solid solutions is *two-gap s-wave superconductivity* with a strong coupling upper band ($\lambda_{e-ph}$ ~ 1.0, $\Delta_1$ ~ 14 K and $2\Delta_1/k_BT_c \approx 4.8$) and weak coupling lower band ($\lambda_{e-ph}$ ~ 0.1-0.4, $\Delta_2$ ~ 6-8 K and $2\Delta_2/k_BT_c$ ~ 2 ), and with strong impurity scattering and interband coupling. At the same time we cannot exclude *anisotropic single-band superconductivity* with a strong anisotropic pair-breaking effect produced by dynamic charge stripes along <110> directions. Moreover, a pseudogap state is detected in $Lu_xZr_{1-x}B_{12}$ with the $\Delta_{ps-gap}$ values varying in the range 60-110 K. Also discussed is the mechanism responsible for the unique enhanced surface superconductivity in $ZrB_{12}$.

### Acknowledgments

This work was supported in part by the Russian Science Foundation, Projects No. 17-12-01426 and No. 21-12-00186 and was performed using the equipment of the Shared Facility Center for Studies of HTS and Other Strongly Correlated Materials, Lebedev Physical Institute, the Russian Academy of Sciences, and of the Center of Excellence, Slovak Academy of Sciences. The work of K.F. and S.G. is supported by the Slovak agencies VEGA (Grant No. 2/0032/20) and APVV (Grant No. 17–0020). The authors are grateful to V.N. Krasnorussky for experimental assistance.

magnetization curves in the superconducting state (Fig. S15), temperature dependences of GLM parameters (Fig. S16) and the scaling of thermodynamic and upper critical fields (Fig.S17).

**Figure captions**

**Fig.1.** (a) Crystal structure of $R$B$_{12}$. The color plane shows the distribution of the electron density in dynamic charge stripes (green bands) along with [110] direction as observed in [43] at T = 50 K. Panels (b)-(c) demonstrate the electron density changes between T = 293 K (b) and 50 K (c) as deduced from x-rays diffraction experiment by the maximal entropy method [43]. (d) Fragment of RB$_{12}$ crystal structure composed from two truncated cuboctahedrons B$_{24}$ centered by Lu/Zr ions. (e) Schematic view of two double-well potentials in vicinity of Lu/Zr ions with oscillations of metallic ions from their central positions inside the B$_{24}$ cubooctahedra. The barrier height in the double-well potential (the pseudogap) is about equal to the cage-glass transition temperature $\Delta_{ps\text{-}gap} \cong T^*$. (f) The lattice constant $a(x)$ variation in Lu$_x$Zr$_{1-x}$B$_{12}$.

**Fig.2.** (a)-(b) The temperature dependencies of resistivity $\rho(T)$ of several Lu$_x$Zr$_{1-x}$B$_{12}$ crystals. (c) The large scale of $\rho(T)$ near $T_c$ demonstrates the superconducting transition for Zr-rich samples of Lu$_x$Zr$_{1-x}$B$_{12}$. (d) The diamagnetic response measured in field cooled (1-6 Oe) MPMS experiments for $x < 0.46$ Lu composition and in AC magnetic susceptibility studies in $^3$He-$^4$He minifridge.

**Fig.3.** (a) Residual resistivity $\rho_0(x)$ and slop $\alpha(x)$ of the $\rho(T, x) = \rho_0(x)+\alpha(x)T$ dependence as estimated in temperature range 80-300 K. (b) Concentration dependence of the GLM parameters $\kappa_1(0)$ and $\kappa_2(0)$ in Lu$_x$Zr$_{1-x}$B$_{12}$. Inset in panel (b) shows a phase diagram fragment in the superconducting state of ZrB$_{12}$ where type-*I*, type-*II/1*, and type-*II/2* phases are shown (see also text). Vertical and horizontal dashed lines show the Lu percolation threshold at $x_c \sim 0.23$ and the critical value $\kappa_c = 1/\sqrt{2}$ on the boundary of type-*I* and type-*II* superconductivity, correspondingly.

**Fig.4.** Field dependencies of magnetoresistance $\Delta\rho/\rho(H,T = 4.2$ K$)$ (panels (a)-(b)) and Hall coefficient $R_H(H,T=4.2$ K$)$ (c) measured in Lu$_x$Zr$_{1-x}$B$_{12}$. (d) Temperature dependencies of drift $\mu_D(T)$ and Hall $\mu_H = R_H/\rho(T)$ mobilities of charge carriers measured for several Lu$_x$Zr$_{1-x}$B$_{12}$ crystals at $H=80$ kOe. (e) Concentration dependencies of drift $\mu_D(x, T=4.2K)$ and Hall $\mu_H(x, T=4.2K)$ mobilities and the average relaxation time $\tau(x)$ (see text for more detail). Vertical dashed line shows the Lu percolation threshold at $x_c \sim 0.23$.

**Fig.5.** (a) Zero field heat capacity temperature dependences $C(T)$ (right-bottom) and corresponding $C(T)/T = f(T^2)$ curves measured in magnetic field of about 1 kOe which destroys superconductivity (left-top) of a number of investigated Lu$_x$Zr$_{1-x}$B$_{12}$ single crystals. Panels (b) and (c) highlight the zero-field normalized heat capacity behavior in the superconducting state of (b) Lu$_{0.104}$Zr$_{0.896}$B$_{12}$ and (c) Lu$_{0.17}$Zr$_{0.83}$B$_{12}$. Thick solid and dashed lines show the fits using the two-band and single-band $\alpha$-models (see text), correspondingly. Thin lines present the smaller-gap and larger-gap components.

**Fig.6.** (a) Pseudogap $\Delta_{ps\text{-}gap}(x) \cong T^*(x)$ and superconducting transition temperature $T_c(x)$ changes in Lu$_x$Zr$_{1-x}$B$_{12}$ solid solutions. The color areas show the pseudogap (green) and superconducting (pink) states. (b) Concentration dependencies of the single-band average superconducting gap $\Delta(0)$ (see Eq.(10)) and the two-band gaps $\Delta_1(0)$ and $\Delta_2(0)$ derived within the $\alpha$-model in Lu$_x$Zr$_{1-x}$B$_{12}$. Inset presents the $x$-dependence of the relative weight $n_2(x)$ of the smaller-gap component.

**Fig.7.** Low temperature heat capacity dependencies $C(T, H_0)/T$ of (a) $Lu_{0.104}Zr_{0.896}B_{12}$, (b) $Lu_{0.46}Zr_{0.54}B_{12}$ and (c) $Lu_{0.74}Zr_{0.26}B_{12}$ in small magnetic fields which destroy the superconductivity.

**Fig.8.** Lu content dependencies of (a) the Sommerfeld coefficient $\gamma(x)$ and the normalized concentration of charge carriers $n_e/n_R(x) = 1/(R_H e\, n_R)$, and (b) the jump of heat capacity $\Delta C/T_c(x)$ and the average Fermi velocity $v_F(x)$. Inset in panel (b) shows changes of the electron-phonon interaction constant $\lambda_{e\text{-}ph}(x)$ in $Lu_xZr_{1-x}B_{12}$ (see text).

**Fig.9.** Temperature dependencies of the thermodynamic $H_{cm}(T)$ and upper $H_{c2}(T)$ critical fields, resulting from the heat capacity analysis. SC and N denote the superconducting and normal states. Solid and dashed lines show the fits to highlight $H_{cm}(0)$ and $H_{c2}(0)$ values.

**Fig. 10.** Concentration dependencies of (a) the critical fields $H_{c1}(0)$, $H_{cm}(0)$ and $H_{c2}(0)$, and (b) the coherence length $\xi(0)$, the penetration depth $\lambda(0) = \kappa_{1,2}(0)\cdot\xi(0)$ and the mean free parh of charge carriers $l(x)$.

**Fig.11.** The diamagnetic $M(H, T_0)$ dependencies of $Lu_xZr_{1-x}B_{12}$ samples with $x = 0.04$, $0.1$ and $0.2$ (panels a, b and c, respectively). The procedure usually applied for the extraction of critical fields $H_{c1}$ and $H_{c2}$ is shown in the insets.

**Fig.12.** Temperature dependencies of the normalized field of superconductivity suppression in Lu-rich $Lu_xZr_{1-x}B_{12}$ samples in comparison with the BCS formula Eq.(6) for the thermodynamic critical field and with the empiric relation $h_{c2}(t) = (1-t^2)/(1+t^2)$ $t = T/T_c$ for the upper critical field.

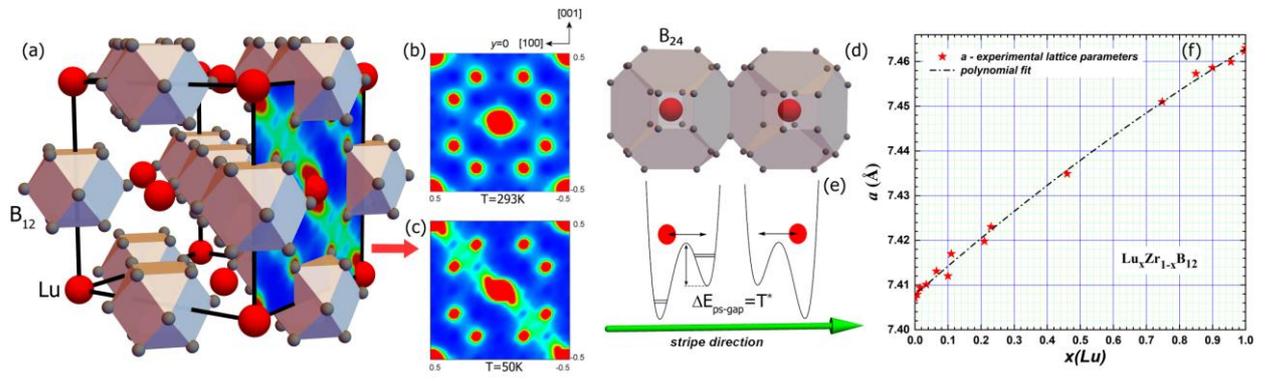

Fig.1.

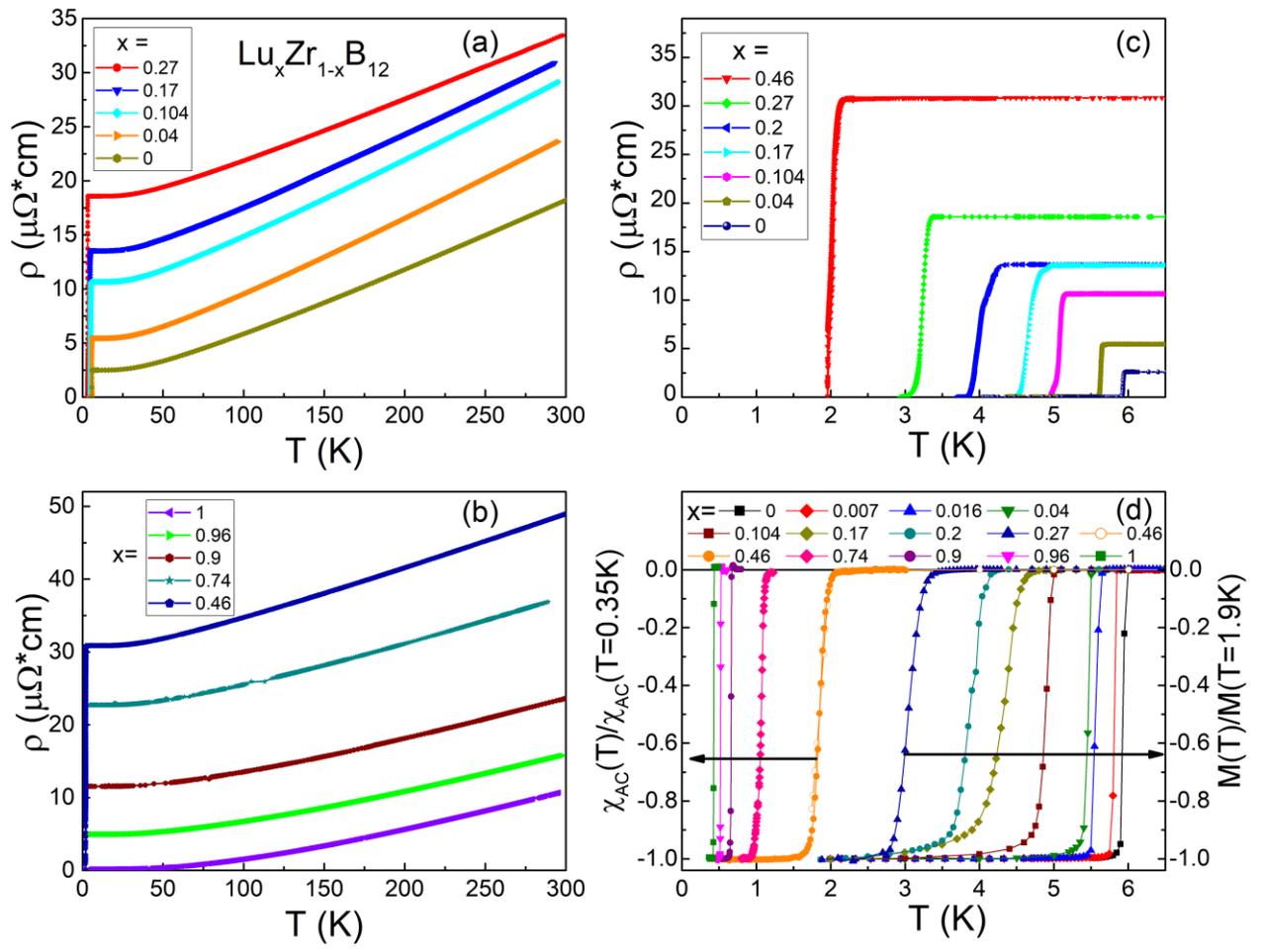

Fig.2.

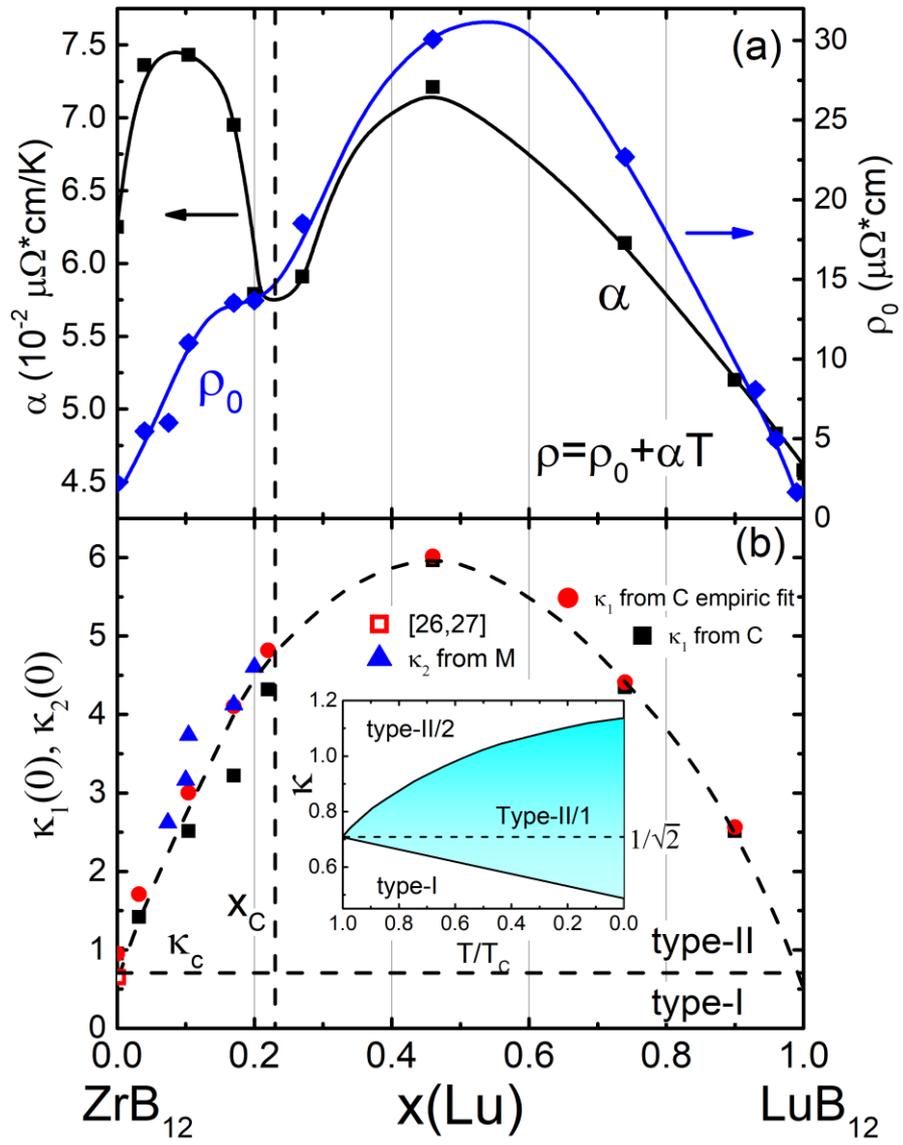

Fig.3.

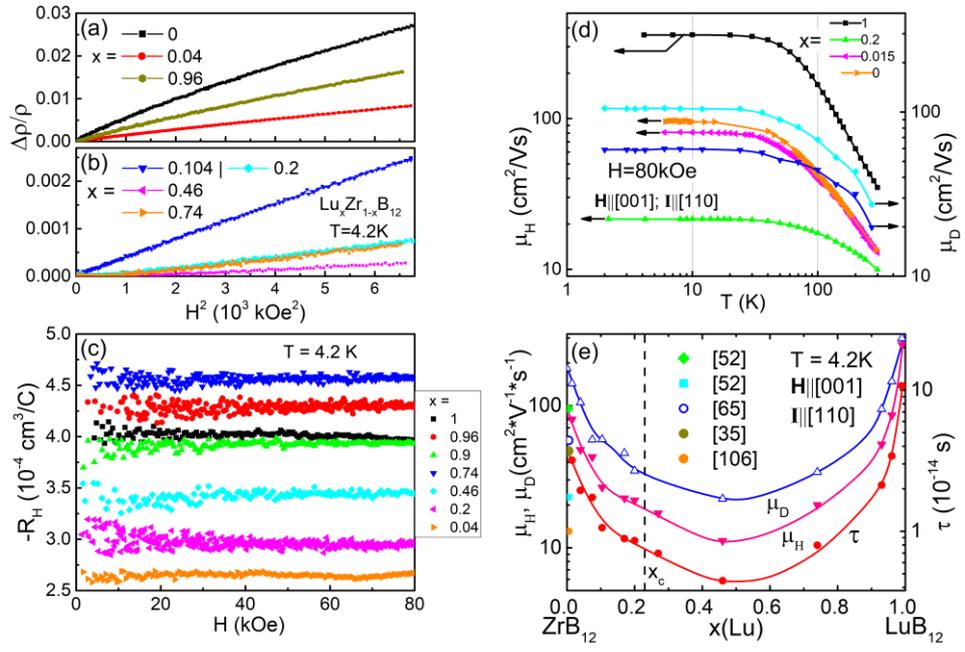

Fig.4.

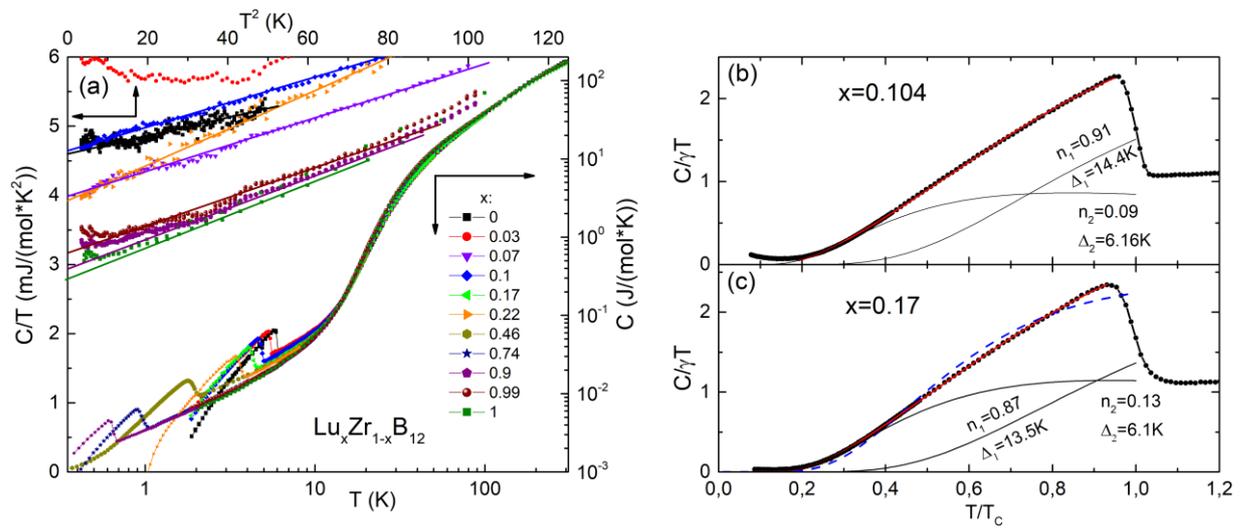

Fig.5.

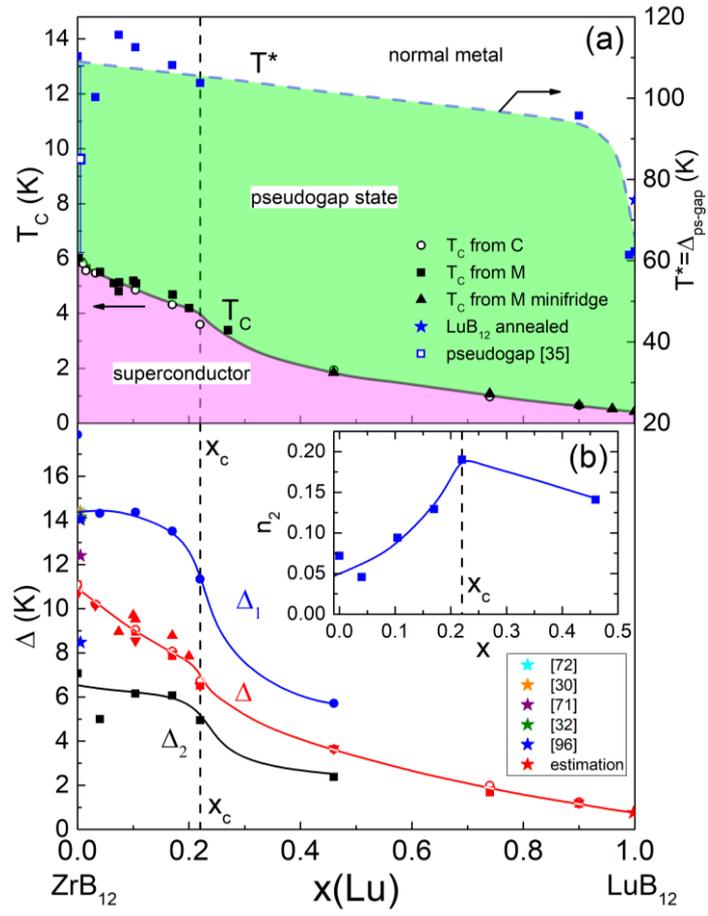

Fig.6.

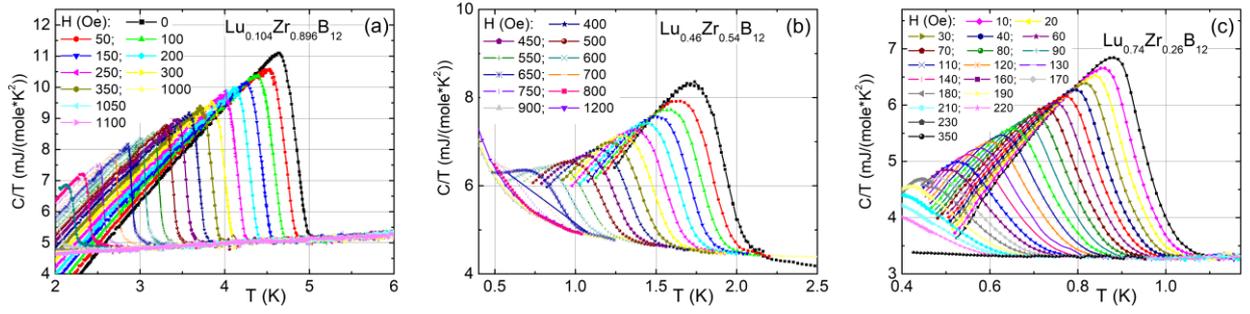

Fig.7.

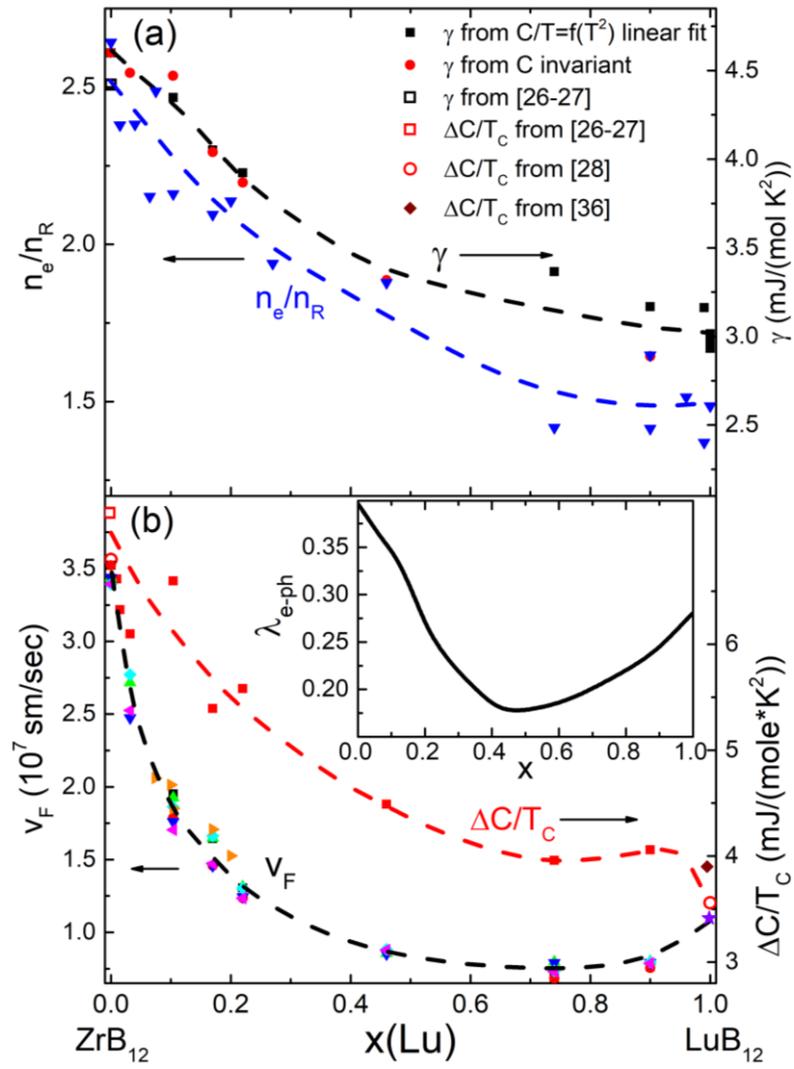

Fig.8.

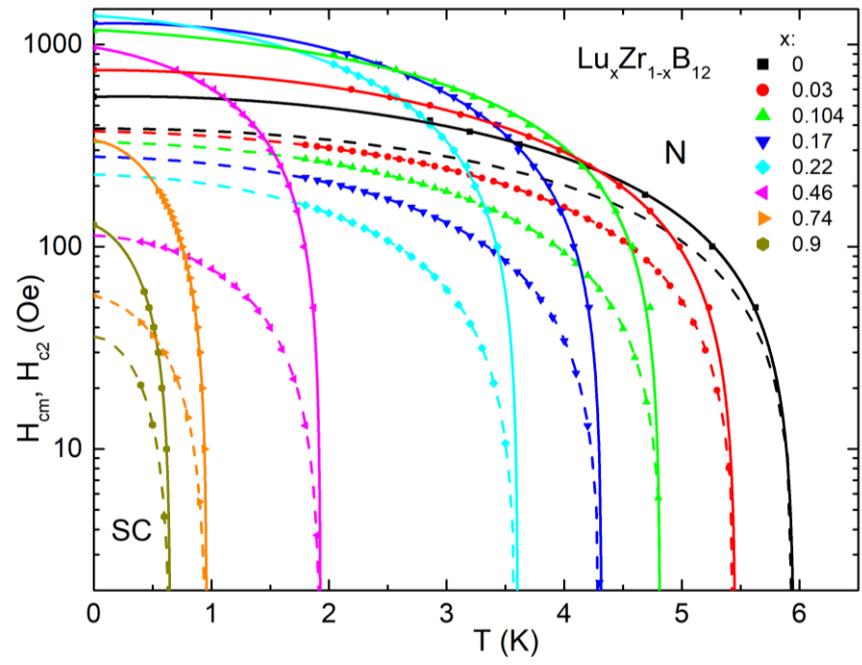

Fig.9.

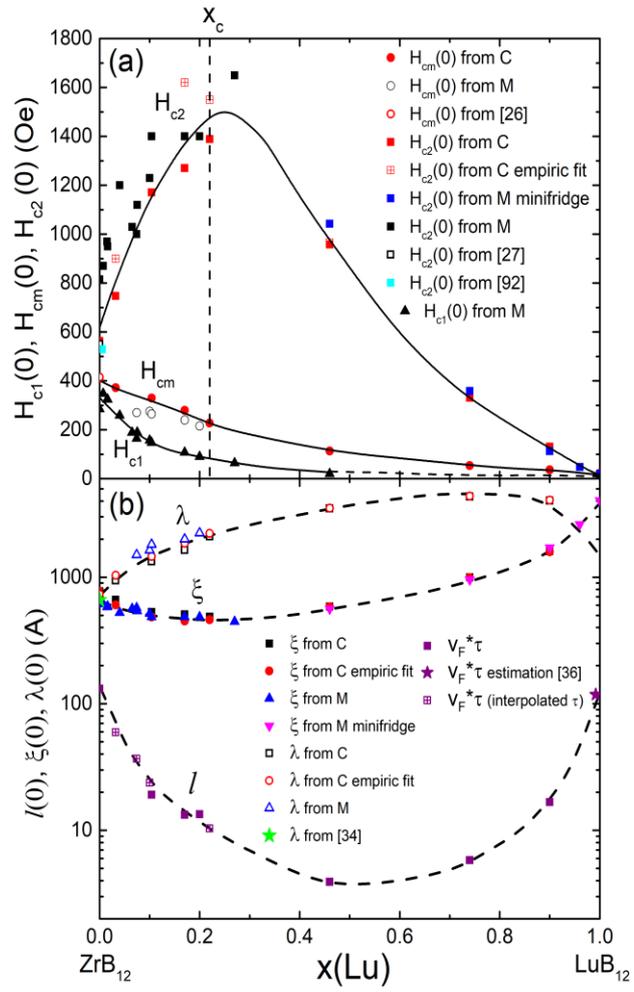

Fig.10.

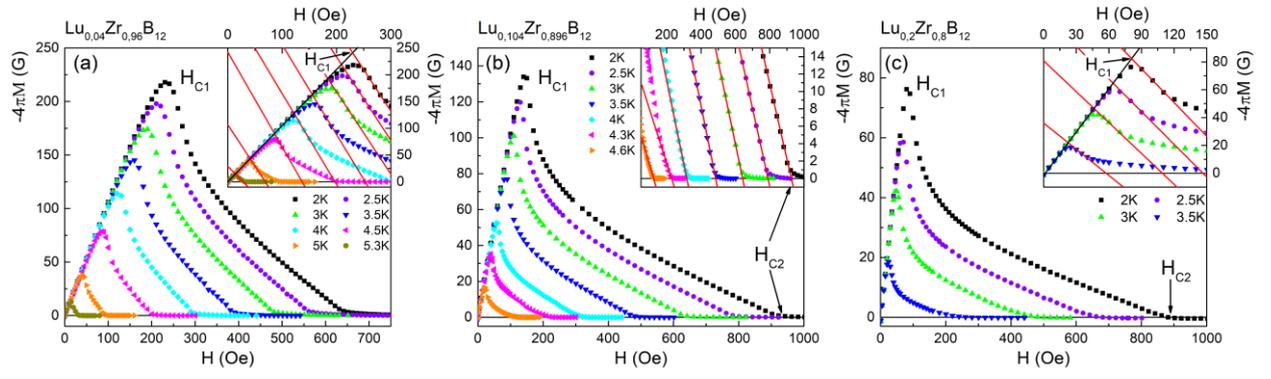

Fig.11.

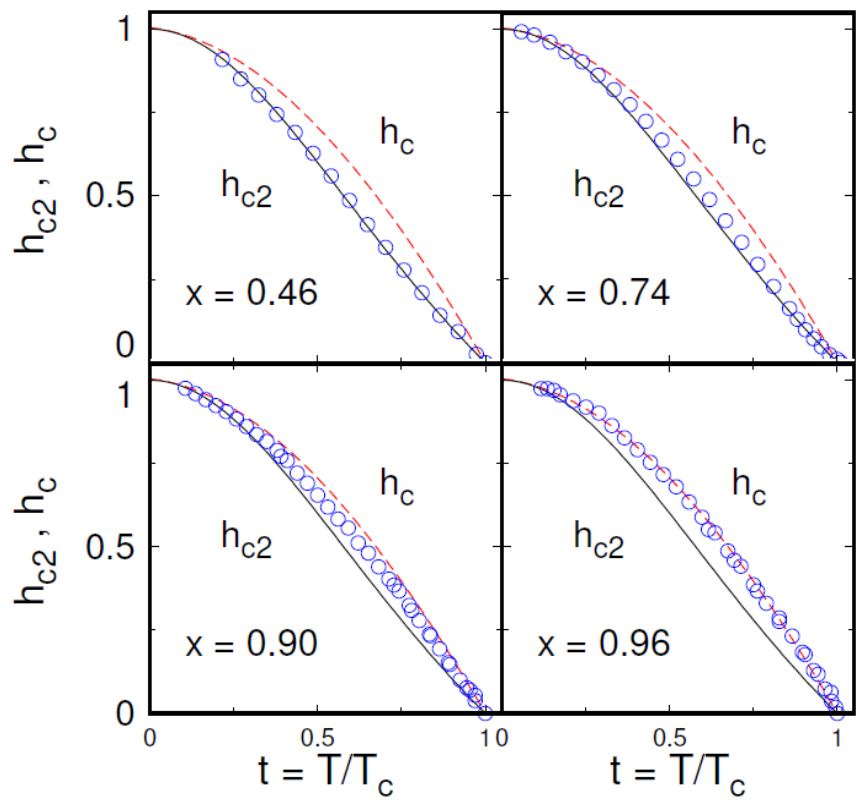

Fig.12.

# Supplementary information to the paper

# Inhomogeneous superconductivity in $Lu_xZr_{1-x}B_{12}$ dodecaborides

# with dynamic charge stripes

A. Azarevich[1], A. Bogach[1], V. Glushkov[1], S. Demishev[1], A. Khoroshilov[1], K. Krasikov[1],

V. Voronov[1], N. Shitsevalova[2], V. Filipov[2], S. Gabáni[3], K. Flachbart[3], A. Kuznetsov[4],

S. Gavrilkin[5], K. Mitsen[5], S.J. Blundell[6], N. Sluchanko[1]

[1]Prokhorov General Physics Institute of the Russian Academy of Sciences, Vavilov str. 38, Moscow 119991, Russia

[2]Frantsevich Institute for Problems of Materials Science, National Academy of Sciences of Ukraine, 03680 Kyiv, Ukraine

[3]Institute of Experimental Physics SAS, Watsonova 47, 04001 Košice, Slovakia

[4]National Research Nuclear University MEPhI, 31, Kashirskoe Shosse, 115409 Moscow, Russia

[5]Lebedev Physical Institute of RAS, 53 Leninskiy Avenue, 119991 Moscow, Russia

[6]Department of Physics, University of Oxford, Clarendon Laboratory, Parks Road, Oxford OX1 3PU, United Kingdom

1. **The crystal growth of $Lu_xZr_{1-x}B_{12}$ substitutional solid solutions**

The single crystals of $Lu_xZr_{1-x}B_{12}$ solid solutions were grown using crucible-free inductive floating zone technique with "Crystal-111A" special unit at the Institute for Problems of Materials Science, National Academy of Sciences of Ukraine (Kyiv). Source $Lu_xZr_{1-x}B_{12}$ sintered rods for melting were prepared from corresponding powders synthesized by a reduction of mixture of lutetium and zirconium metal oxides taken in corresponding ratio in presence of boron at 1900 K in vacuum according to the routine solid-state reaction:

½$x$Lu$_2$O$_3$ + (1-$x$)ZrO$_2$ + ($n$+3.5-2$x$)B → Lu$_x$Zr$_{1-x}$B$_n$ + (3.5-2$x$)BO↑      (0≤ $x$ ≤ 1.0; 12.5 ≤ $n$ ≤ 13.8).

The basic substance content in initial Lu$_2$O$_3$ and ZrO$_2$ oxides was 99.998 and 99.9 wt. % respectively, and the content in initial amorphous boron was higher than 99.5 wt. %. The highly volatile impurities presented in boron were removed during the synthesis procedure and, after that, in the zone melting process.

The source powders were synthesized with boron excess taking into account the specific nature of ZrB$_{12}$ melting behavior. ZrB$_{12}$ is stable only in a narrow temperature range, (1696 ÷ 2082)$^0$C, and melts peritectically with the formation of ZrB$_2$ and a boron-rich melt (Fig. S1). Zirconium-rich Lu$_x$Zr$_{1-x}$B$_{12}$ solid solutions melt in a similar way. Introducing an excess of boron into initial compositions enriched the molten zone by boron that allowed to reduce the temperature of the melt, to avoid precipitation of the Lu$_x$Zr$_{1-x}$B$_2$ phase and to improve the crystal quality. The amount of boron has to be selected empirically for each solid solution composition in order to avoid crystallization of Lu$_x$Zr$_{1-x}$B$_{51}$ solid solutions as an impurity phase.

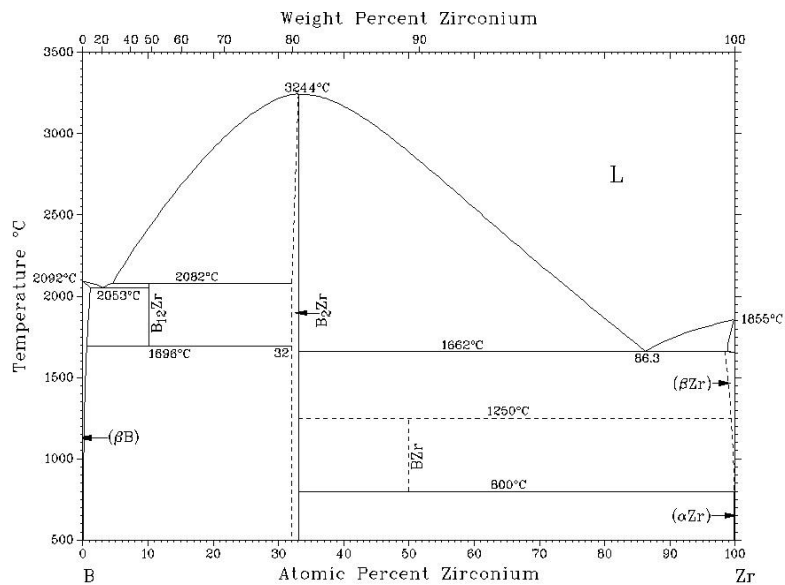

**Fig. S1.** B - Zr phase diagram [S1].

X-ray phase analysis of synthesized source powders has shown that all powders are two-phase products, containing diboride ($Me$B$_2$: AlB$_2$ struct. type) and dodecaboride ($Me$B$_{12}$: UB$_{12}$ struct. type) phases (Fig. S2). Besides, the presence of a boron-rich phase of the ZrB$_{51}$ type [S2] in synthesized powders cannot be ruled out, taking into account the excess of boron in the source charge, the large concentration of diboride in synthesized products, and small intensity of X-ray reflections from this substance on the background of diboride and dodecaboride phase's reflections. So, the synthesized powders were considered as a ligature for growth of Lu$_x$Zr$_{1-x}$B$_{12}$ single crystals with a traveling solvent floating zone technique.

Taking into account the Zr - B phase diagram (Fig. S1) and difference in Lu and Zr metals distribution coefficients the main task was to avoid the precipitation of impurity phases and to maintain a constant Lu/Zr ratio along the growth axis and in lateral cross-section of the crystal. To solve this problem we optimized individually for each Lu$_x$Zr$_{1-x}$B$_n$ composition such technological parameters as the inert gas pressure (highly pure Ar, $P \leq 1.0$ MPa), the crystallization rate (0.15…0.30 mm/min), the rate of rotation of feed and seed rods (0…10 rpm). Laue back scattering patterns were used for primary analysis of crystal structure quality, and the technology has been worked out in such a way in order to avoid the splitting of point reflections in the Laue pattern that correspond to the absence of block misorientation within a few tenths of a degree (procedure resolution). A specific feature of the zone melting of Lu$_x$Zr$_{1-x}$B$_n$ compositions is the crystallization of $Me$B$_2$ and $Me$B$_{51}$ as impurities phases at the initial moment of alloying of the seed and sintered rod. The stable growth of the dodecaboride phase begins after several mm of crystal growth. As a result, single crystals of all nominal compositions Lu$_x$Zr$_{1-x}$B$_n$ ($0 \leq x \leq 1$) were grown with diameters of 5–6 mm and a length of about 50 mm.

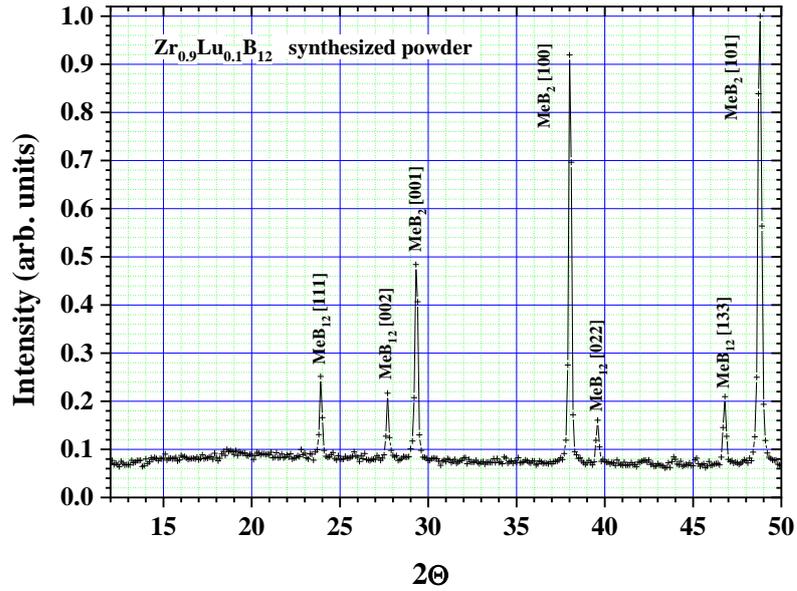

**Fig. S2.** X-ray diffraction pattern of the synthesized powder for the $Lu_{0.1}Zr_{0.9}B_{13.5}$ nominal composition in Co $K_\alpha$ radiation with Fe filter.

The ratio Lu/Zr was estimated for all $Lu_xZr_{1-x}B_n$ single crystals using a scanning electron microscope equipped with a system of the energy dispersion microprobe (JEOL JXA-8200 EPMA; electron probe size 1 $\mu m^2$). The measurements were carried out at several points of lateral cross section (periphery $r = 1$, middle $r = ½$, centre $r = 0$) and on both sides of the single crystal rods. Individual binary borides ($ZrB_2$, $ZrB_{12}$, $ZrB_{51}$ and $LuB_{12}$) were used as reference samples. The accuracy of the microprobe analysis is several hundred ppm (according to the registration certificate). The real compositions of single crystals differ from the nominal ones and are slightly different along the crystal, but remain constant within every lateral cross section.

The main part of the crystal rods consisted of a single crystal core that was free of domain boundaries, surrounded by a thin polycrystalline ring (Fig. S3). Electron microscopy showed that the initial part of the crystal (about 3 mm along the growth axis) adjacent to the melting beginning contained inclusions of both $MeB_2$ and $MeB_{51}$ phases (Fig. S4a). The periphery of the crystal final part (lateral cross section) included the impurity phase of $ZrB_{51}$- type, in particular for the crystal with nominal composition $Lu_{0.1}Zr_{0.9}B_{13.5}$ the composition of the impurity phase was $Lu_{0.016}Zr_{0.984}B_{51.32}$ (Fig. S4b).

In all cases, the single crystalline samples for the present study were cut from the central part of oriented polished plates (discs, $h \approx 0.6 \div 0.8$ mm), free from impurity phases. X-ray powder analysis of crushed $Lu_xZr_{1-x}B_{12}$ single crystals revealed reflections of the $UB_{12}$ structure type only (Fig. S5). The real compositions of single crystals differ from the nominal ones (Fig. S6) and are slightly different along the crystal, but remain constant within every lateral cross section.

For example, the results of microprobe analyses of a single crystal grown from a source sintered rod with nominal composition of $Lu_{0.1}Zr_{0.9}B_{13.5}$ indicated that the actual Lu content changes from $x = 0.065(1)$ at the zone beginning to $x = 0.074(1)$ at the zone end. The Laue back reflection pattern of this crystal is shown in Fig. S7.

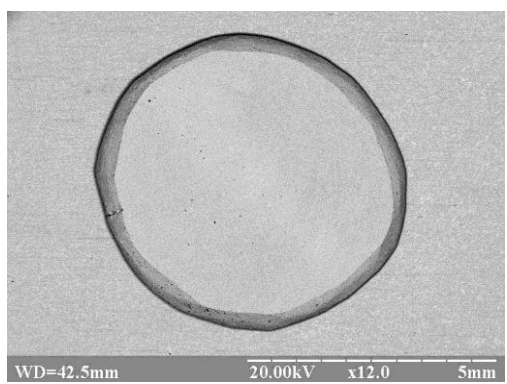

**Fig. S3.** General view of the lateral cross-section of the central part of the single crystal with nominal composition $Lu_{0.1}Zr_{0.9}B_{13.5}$.

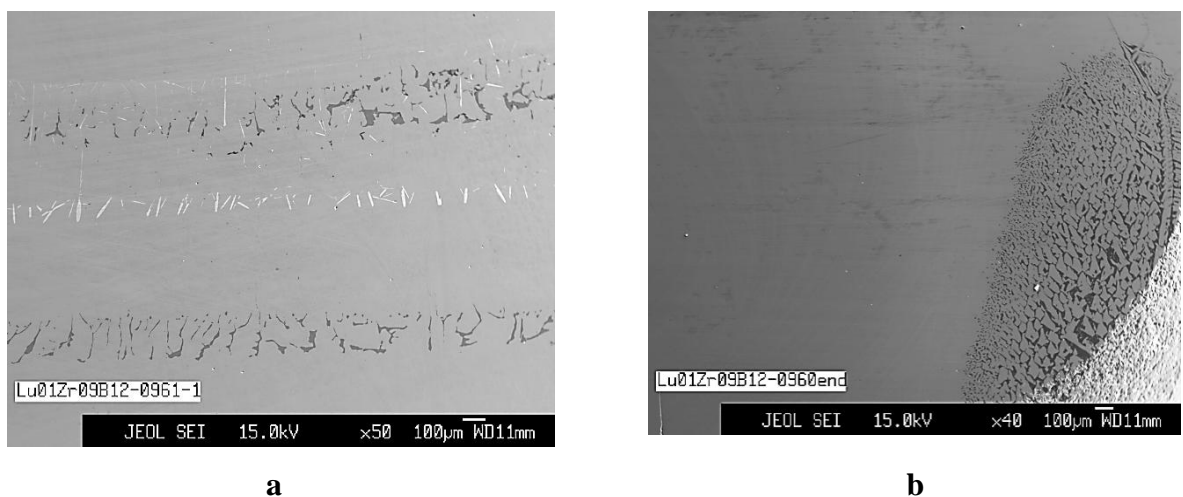

**Fig. S4.** Lateral cross-sections of the (**a**) initial and (**b**) main parts of a single crystal with nominal composition $Lu_{0.1}Zr_{0.9}B_{13.5}$. The light and dark inclusions in (**a**) are $MeB_2$ and $MeB_{51}$ phases, respectively; dark inclusions in (**b**) at the single crystal periphery is $MeB_{51}$ phase with a real $Zr_{0.984}Lu_{0.016}B_{51.32}$ composition.

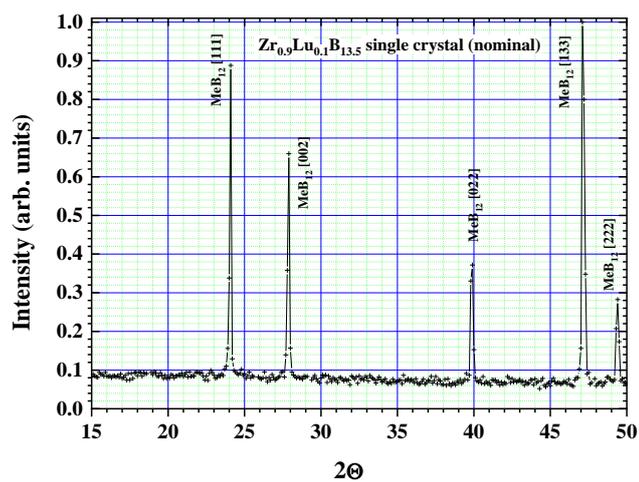

**Fig. S5.** X-ray diffraction pattern of the $Zr_{0.9}Lu_{0.1}B_{13.5}$ single crystal in Co $K_\alpha$ radiation with Fe filter.

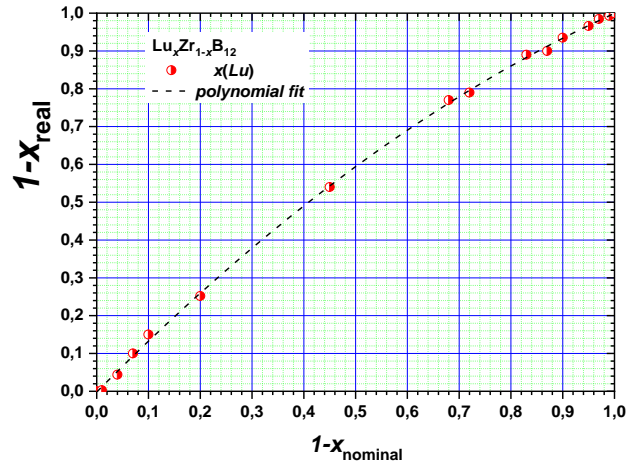

**Fig. S6.** Real Zr/Lu ratios versus nominal ones for $Lu_xZr_{1-x}B_{12}$ single crystals.

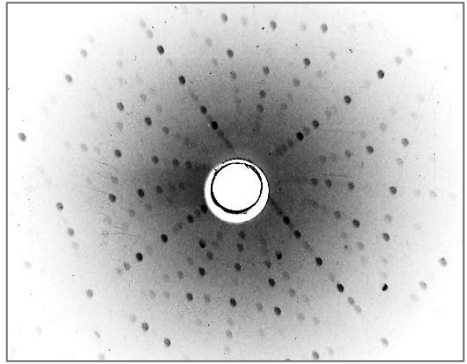

**Fig. S7.** X-ray Laue pattern from a lateral cross section of as-grown single crystal with nominal composition $Lu_{0.1}Zr_{0.9}B_{13.5}$ grown with a [100] oriented seed. Deviation of the growth direction from [100] is about 3 degree.

An optical spectral analysis was used additionally to estimate the impurities in crystals. The total concentration of impurities (except for the rare-earth elements) did not exceed $10^{-2}$ wt. %, rare-earth impurities are determined by the purity of starting lutetium and zirconium oxides.

## 2. Magnetoresistance and Hall coefficient.

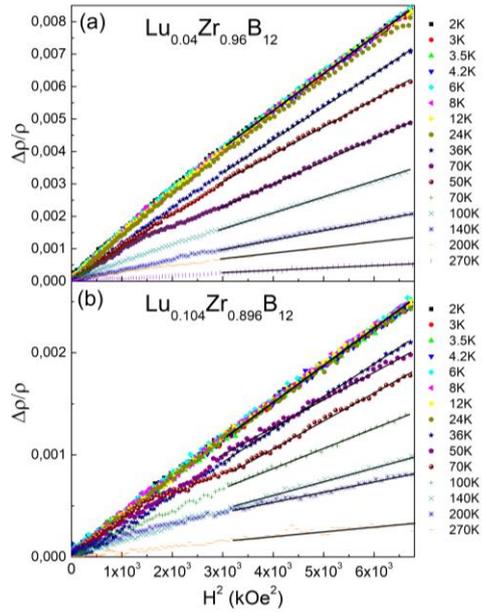

**Fig.S8.** Magnetic field dependencies of magnetoresistance of (a) $Lu_{0.04}Zr_{0.96}B_{12}$ and (b) $Lu_{0.104}Zr_{0.896}B_{12}$ at temperatures in the range 2-270 K. Solid lines show the approximation by the quadratic field dependence $\Delta\rho/\rho=\mu_D H^2$ (see text of paper for more details).

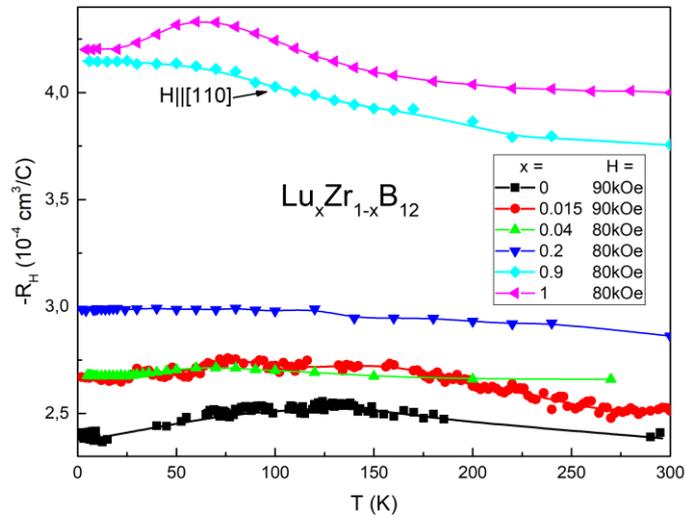

**Fig.S9.** Temperature dependences of the Hall coefficient for $Lu_xZr_{1-x}B_{12}$ samples in strong magnetic field.

### 3. Heat capacity components in Lu$_x$Zr$_{1-x}$B$_{12}$.

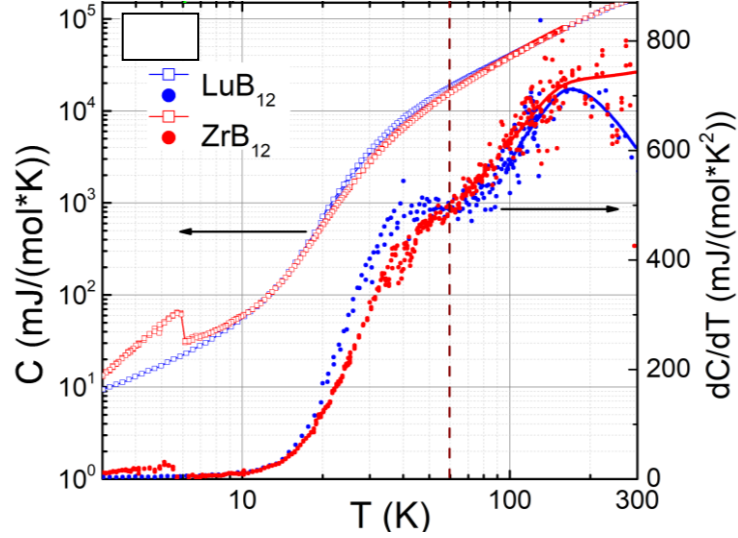

**Fig.S10**. Temperature dependencies of the specific heat and the derivative $dC/dT=f(T)$ for LuB$_{12}$ and ZrB$_{12}$.

To analyze the heat capacity contributions of Lu$_x$Zr$_{1-x}$B$_{12}$ samples, we used an approach similar to that previously applied in [S3-S5]. Contributions from B and Lu/Zr atoms in the vibrational heat capacity are considered in terms of the Debye [Eq. (S1)] and Einstein [Eq. (S2)] models, respectively:

$$\frac{C_D}{T^3} = \frac{9rR}{T_D^3} \int_0^{T_D/T} \frac{x^4 e^{-x} dx}{(1-e^{-x})^2} \tag{S1}$$

$$\frac{C_E}{T^3} = \frac{3R}{T_E^3}\left(\frac{T_E}{T}\right)^5 \frac{e^{-\frac{T_E}{T}}}{(1-e^{-\frac{T_E}{T}})^2}, \tag{S2}$$

where $R$ is gas constant and $r = 12$. The electronic specific heat $C_{el} = \gamma T$ with $\gamma$ and the Debye contribution with Debye temperature $T_D$ (see [S3-S5] for more detail) were used to calculate the difference $C_{ph} = C - C_{el} - C_D = C_E(T) + C_{Sch(i)}(T)$. Following to [S3-S5] we used additionally two Schottky terms $C_{Sch(i)}$ – two-level-systems TLS$_1$ and TLS$_2$ to approximate the low temperature anomalies of heat capacity. It has been argued in [S3-S5] that these two Schottky components:

$$\frac{C_{Sch(i)}}{T^3} = \frac{RN_i}{T^3}\left(\frac{\Delta E_i}{T}\right)^2 \frac{e^{\frac{\Delta E_i}{T}}}{(e^{\frac{\Delta E_i}{T}}+1)^2} \qquad (S3)$$

($i$ = 1, 2 and $N_i$ is the concentration of two-level-systems) are necessary to describe the effect of boron vacancies (TLS$_2$ contribution) and divacancies (TLS$_1$) in the specific heat of $R$B$_{12}$. Indeed, in view of a weak coupling of rare-earth ions in the boron network in combination with a significant number of boron vacancies and other intrinsic defects in the UB$_{12}$ type structure [S6], the formation of double-well potentials (DWP) should be expected at displacements of Lu$^{3+}$/Zr$^{4+}$ ions from central positions in B$_{24}$ cubooctahedra. All the above mentioned specific heat contributions are shown in Fig. S11 in the $C/T^3$ plot together with our experimental data.

According to the approach developed in [S3-S5], the energy of $\Delta E_2/k_B$ = 54–100 K (see Eq. (S3)), which is deduced from the analysis developed here, should be attributed to the barrier height in the DWP.

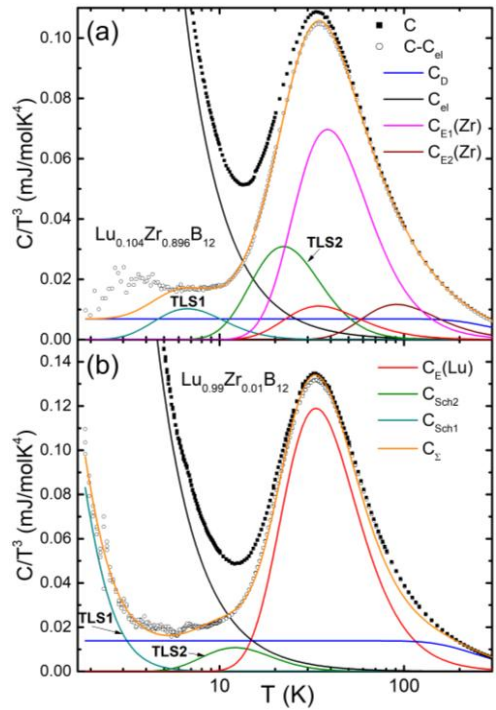

**Fig. S11**. Analysis of the heat capacity components in (a) Lu$_{0.104}$Zr$_{0.896}$B$_{12}$ and (b) Lu$_{0.99}$Zr$_{0.01}$B$_{12}$. $C$, $C_D$, $C_E$(Zr), $C_E$(Lu) $C_{el}$, $C_{Sch1}$ and $C_{Sch2}$ are experimental curves, the Debye, Einstein of Zr ions, Einstein of Lu ions, electronic (Sommerfeld), two Schottky (TLS$_1$ and TLS$_2$) contributions and $C_\Sigma$ is their sum, correspondingly (see [S3-S5] for more detail).

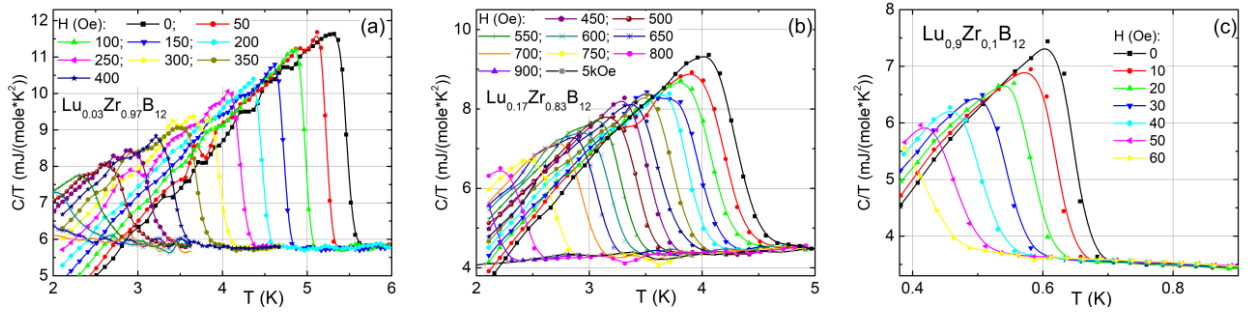

**Fig. S12.** Low temperature heat capacity dependences $C(T, H_0)/T$ of (a) $Lu_{0.03}Zr_{0.97}B_{12}$, (b) $Lu_{0.17}Zr_{0.83}B_{12}$ and (c) $Lu_{0.9}Zr_{0.1}B_{12}$ in small magnetic fields which destroy the superconductivity.

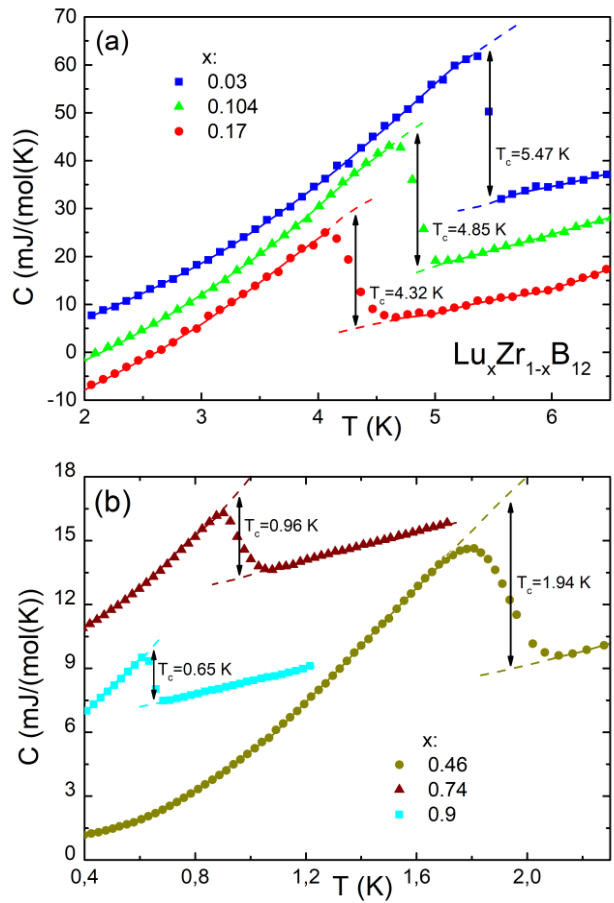

**Fig. S13.** Panels (a) and (b) show the procedure applied to determine the $\Delta C$ jump amplitude near $T_c$ for several samples of $Lu_xZr_{1-x}B_{12}$. The curves are shifted along vertical axis for convenience.

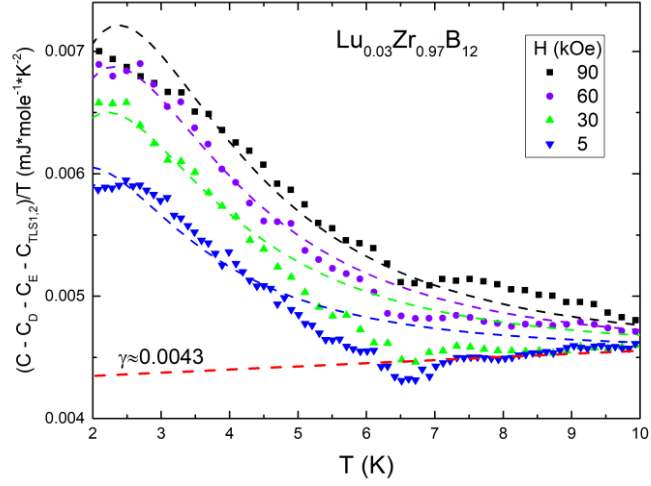

**Fig. S14.** The sum of magnetic and Sommerfeld components in magnetic fields up to 90 kOe. It can be seen clearly, that the field dependent magnetic contribution increases strongly with magnetic field and it prevails the electronic term.

4. **Magnetization, Ginzburg-Landau-Maki parameters and scaling of critical fields.**

Field dependencies of magnetization measured at $T = 2$ K for single crystals of $Lu_xZr_{1-x}B_{12}$ are presented in Fig. S15. For clarity, only $M(H)$ curves obtained in sweeping up field are shown in the top panel. At low fields well below $H_{c1}$ all crystals demonstrate a linear increase of magnetization with a slope corresponding to the total (about 100%) Meissner state. At high fields the magnetization decreases to zero at upper critical field $H_{c2}$ due to suppression of superconductivity. The analysis of these curves has revealed both non-linear and linear $M(H)$ dependences below $H_{c2}$. The non-linear behavior, found in samples with low ($x < 0.065$) and high ($x \geq 0.27$) lutetium content, we attribute to inhomogeneous superconductivity. The linear $M(H)$ behavior below $H_{c2}$ in crystals with $0.074 \leq x \leq 0.23$ allowed us to obtain the GLM parameter from the slope of magnetization curves. Hysteresis on the magnetization curves are shown in the bottom panel of Fig. S15 for some crystals. A decrease of the hysteresis loop with increase of lutetium content was found for $Lu_xZr_{1-x}B_{12}$. For example, the $M(H)$ dependence of the homogenous crystal with $x = 0.075$ remains hysteretic practically up to suppression of superconductivity, while in the inhomogeneous sample with $x = 0.46$ it becomes irreversible only in fields twice lower than $H_{c2}$. We suppose that pinning weakening is caused by the decrease of condensation energy of the superconducting state.

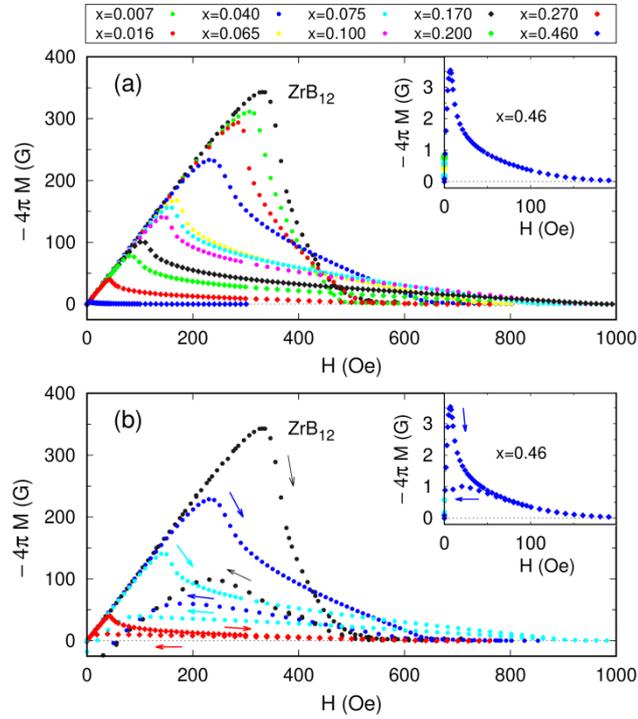

**Fig. S15.** Magnetization curves obtained at $T = 2$ K for $Lu_xZr_{1-x}B_{12}$ single crystals. Only $M(H)$ dependences measured in sweeping up magnetic field experiments are presented in the top panel for clarity. Hysteresis of magnetization is shown in the bottom panel for several crystals.

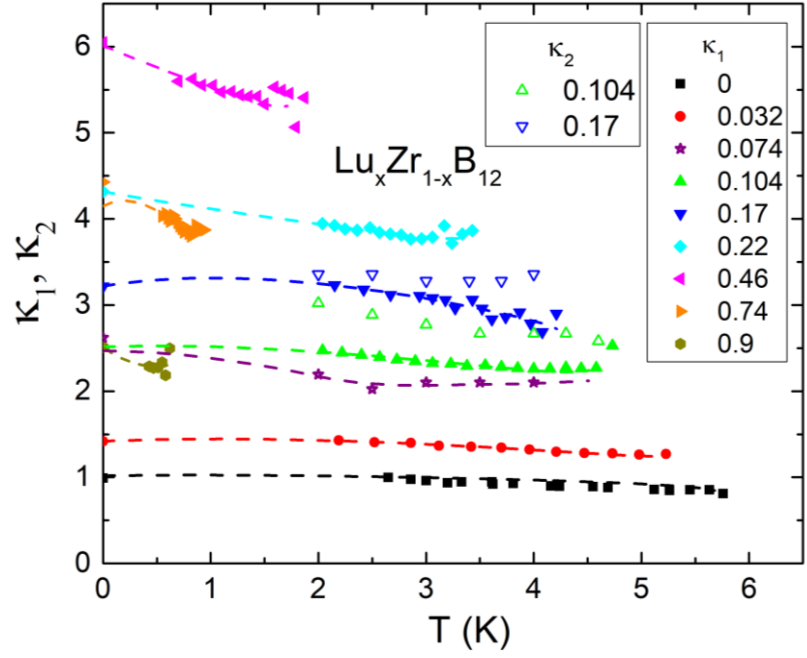

**Fig. S16**. Temperature dependences of the Ginzburg-Landau-Maki parameters in $Lu_xZr_{1-x}B_{12}$ crystals obtained from specific heat ($\kappa_1$) and magnetization ($\kappa_2$) measurements.

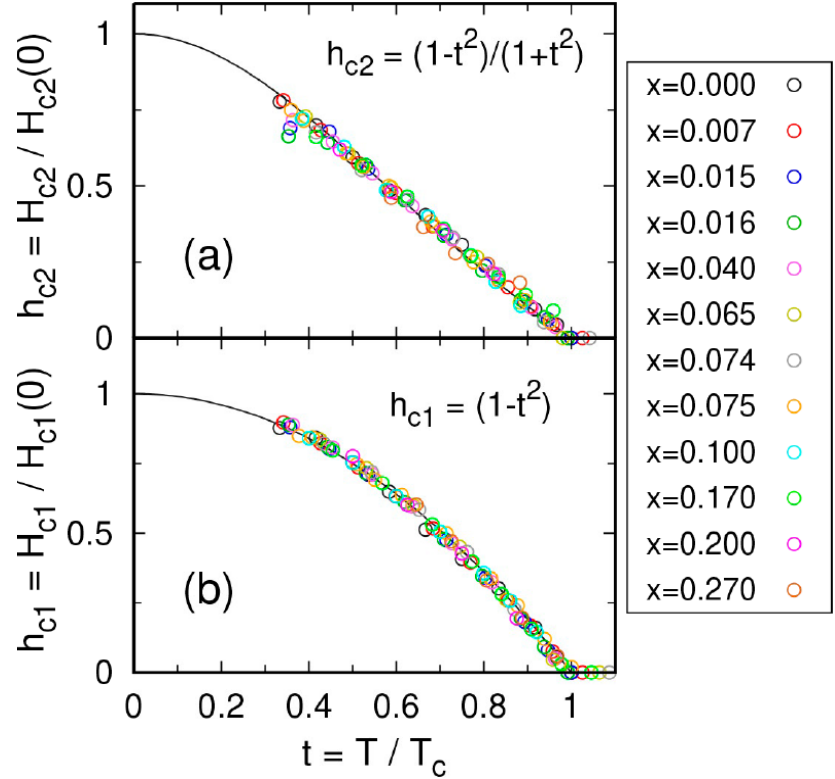

**Fig. S17.** Scaling behavior of the normalized critical fields $h_{c1} = H_{c1}/H_{c1}(0)$ vs $(T/T_c)$ and $h_{c2} = H_{c2}/H_{c2}(0)$ vs $(T/T_c)$ for Zr-rich crystals of $Lu_xZr_{1-x}B_{12}$. Solid lines in panels (a) and (b) show the phenomenological and BCS approximations, correspondingly.